\documentclass[longauth]{aa}
\usepackage[varg]{txfonts}
\usepackage{natbib}
\bibpunct{(}{)}{;}{a}{}{,}
\usepackage{color}
\usepackage{supertabular}

\def\as {\ifmmode {\rlap.}$\,$''$\,$\! \else ${\rlap.}$\,$''$\,$\!$\fi}

\begin{document}

\title{Chemical complexity in high-mass star formation:}

\subtitle{An observational and modeling case study of the AFGL\,2591 VLA\,3 hot core}

 \author{
C.~Gieser\inst{1,2}\and
D.~Semenov\inst{1,3}\and
H.~Beuther\inst{1} \and
A.~Ahmadi\inst{1,2}\and
J.C.~Mottram\inst{1}\and
Th.~Henning\inst{1}\and
M.~Beltran\inst{4}\and
L.T.~Maud\inst{5}\and
F.~Bosco\inst{1,2}\and
S.~Leurini\inst{6}\and
T.~Peters\inst{7}\and
P.~Klaassen\inst{8}\and
R.~Kuiper\inst{9}\and
S.~Feng\inst{10,11}\and
J.S.~Urquhart\inst{12}\and
L.~Moscadelli\inst{4}\and
T.~Csengeri\inst{13} \and
S.~Lumsden\inst{14}\and
J.M.~Winters\inst{15}\and
S.~Suri\inst{1}\and
Q.~Zhang\inst{16}\and
R.~Pudritz\inst{17}\and
A.~Palau\inst{18}\and
K.M.~Menten\inst{13}\and
R.~Galvan-Madrid\inst{18}\and
F.~Wyrowski\inst{13}\and
P.~Schilke\inst{19}\and
\'A.~S\'anchez-Monge\inst{19}\and
H.~Linz\inst{1}\and
K.G.~Johnston\inst{14}\and
I. Jim\'enez-Serra\inst{20}\and
S.~Longmore\inst{21}\and
T.~M{\"o}ller\inst{19}
}
\institute{$^1$ Max Planck Institute for Astronomy, K\"onigstuhl 17,69117 Heidelberg, Germany, \email{gieser@mpia.de}\\
$^2$ Fellow of the International Max Planck Research School for Astronomy and Cosmic Physics at the University of Heidelberg (IMPRS-HD)
$^3$ Department of Chemistry, Ludwig Maximilian University, Butenandtstr. 5-13, 81377 Munich, Germany\\
$^4$ INAF, Osservatorio Astrofisico di Arcetri, Largo E. Fermi 5, I-50125 Firenze, Italy\\
$^5$ European Southern Observatory, Karl-Schwarzschild-Str. 2, D-85748 Garching, Germany\\
$^6$ INAF, Osservatorio Astronomico di Cagliari, Via della Scienza 5, I-09047, Selargius (CA), Italy\\
$^7$ Max-Planck-Institut f\"{u}r Astrophysik, Karl-Schwarzschild-Str. 1, D-85748 Garching, Germany\\
$^8$ UK Astronomy Technology Centre, Royal Observatory Edinburgh, Blackford Hill, Edinburgh EH9 3HJ, UK\\
$^9$ Institute of Astronomy and Astrophysics, University of T\"ubingen, Auf der Morgenstelle 10, 72076, T\"ubingen, Germany\\
$^{10}$ National Astronomical Observatory of China, Datun Road 20, Chaoyang, Beijing, 100012, P. R. China\\
$^{11}$ CAS Key Laboratory of FAST, NAOC, Chinese Academy of Sciences\\
$^{12}$ Centre for Astrophysics and Planetary Science, University of Kent, Canterbury, CT2 7NH, UK\\
$^{13}$ Max Planck Institut for Radioastronomie, Auf dem H\"ugel 69, 53121 Bonn, Germany\\
$^{14}$School of Physics and Astronomy, University of Leeds, Leeds LS2 9JT, United Kingdom\\
$^{15}$ IRAM, 300 rue de la Piscine, Domaine Universitaire de Grenoble, 38406 St.-Martin-d’H\`eres, France\\
$^{16}$ Center for Astrophysics $|$ Harvard \& Smithsonian, 60 Garden Street, Cambridge, MA 02138, USA\\
$^{17}$ Department of Physics and Astronomy, McMaster University, 1280 Main St. W, Hamilton, ON L8S 4M1, Canada\\
$^{18}$ Instituto de Radioastronom\'ia y Astrof\'isica, Universidad Nacional Aut\'onoma de M\'exico, 58090 Morelia, Michoac\'an, M\'exico\\ 
$^{19}$ I. Physikalisches Institut, Universit\"at zu K\"oln, Z\"ulpicher Str. 77, D-50937, K\"oln, Germany\\
$^{20}$ Centro de Astrobiolog\'ia (CSIC, INTA), Ctra. de Ajalvir, km. 4, Torrej\'on de Ardoz, 28850 Madrid, Spain \\
$^{21}$ Astrophysics Research Institute, Liverpool John Moores University, Liverpool, L3 5RF, UK\\
}
\offprints{C. Gieser, \email{gieser@mpia.de}}

\date{Received / Accepted }

\abstract {} {In order to understand the observed molecular diversity in high-mass star-forming regions, we have to determine the underlying physical and chemical structure of those regions at high angular resolution and over a range of evolutionary stages.} {We present a detailed observational and modeling study of the hot core VLA\,3 in the high-mass star-forming region AFGL\,2591, which is a target region of the NOrthern Extended Millimeter Array (NOEMA) large program CORE. Using NOEMA observations at 1.37\,mm with an angular resolution of $\sim$0\as42 ($1\,400$\,au at 3.33\,kpc), we derived the physical and chemical structure of the source. We modeled the observed molecular abundances with the chemical evolution code {\tt MUSCLE} (MUlti Stage ChemicaL codE).} {With the kinetic temperature tracers CH$_3$CN and H$_2$CO we observe a temperature distribution with a power-law index of $q = 0.41\pm0.08$. Using the visibilities of the continuum emission we derive a density structure with a power-law index of $p = 1.7\pm0.1$. The hot core spectra reveal high molecular abundances and a rich diversity in complex molecules. The majority of the molecules have an asymmetric spatial distribution around the forming protostar(s), which indicates a complex physical structure on scales $< 1\,400$\,au. Using {\tt MUSCLE}, we are able to explain the observed molecular abundance of 10 out of 14 modeled species at an estimated hot core chemical age of $\sim$21\,100\,years. In contrast to the observational analysis, our chemical modeling predicts a lower density power-law index of $p < 1.4$. Reasons for this discrepancy are discussed.} {Combining high spatial resolution observations with detailed chemical modeling allows us to derive a concise picture of the physical and chemical structure of the famous AFGL\,2591 hot core. The next steps are to conduct a similar analysis for the whole CORE sample, and then use this analysis to constrain the chemical diversity in high-mass star formation to a much greater depth.}


\maketitle 

\section{Introduction} \label{section:introduction}

\begin{figure}[hbt]
\resizebox{\hsize}{!}{\includegraphics{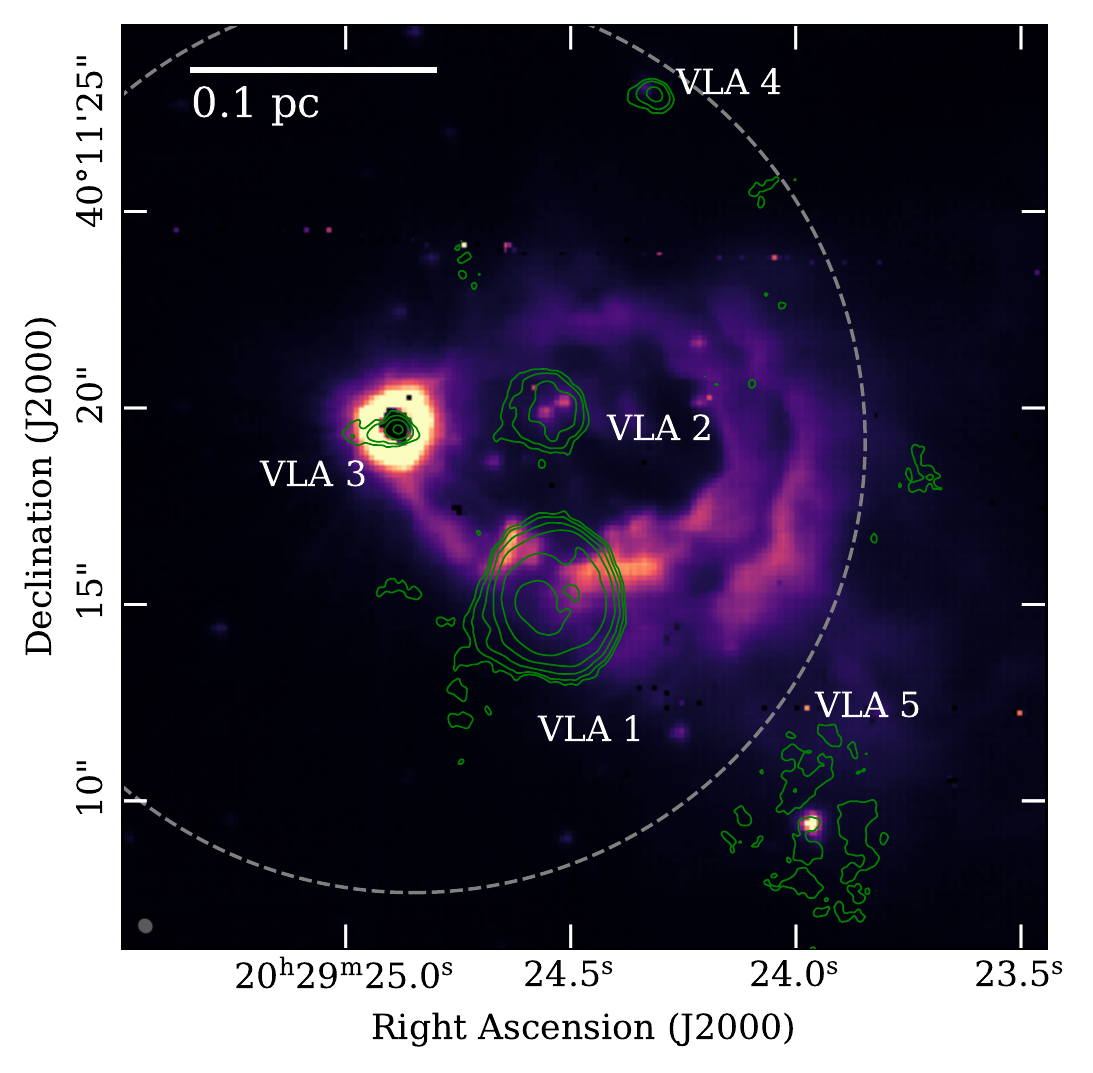}}
\caption{Overview of the AFGL\,2591 star-forming region. The background color map shows an infrared image obtained with Gemini in the $K'$ band. The green contours show the 3.6\,cm radio continuum emission taken from \citet{Johnston2013}. The contours are drawn at levels of 3, 5, 10, 20, 40, 80, and 160$\sigma$ ($\sigma = 0.03$\,mJy\,beam$^{-1}$). The VLA beam size (0\as43$\times$0\as40) is shown in the lower left corner. The gray dashed circle indicates the primary beam of our NOEMA observations (23$''$).}
\label{Fig:AFGLRegion}
\end{figure}

The formation of the most massive stars is an active field of research \citep[for reviews, see, e.g.,][]{Beuther2007, Bonnell2007, Zinnecker2007, Smith2009, Tan2014, Schilke2015, Motte2018}. Since high-mass star-forming regions (HMSFRs) are less abundant than their low-mass counterparts and are typically located at large distances ($> 1$\,kpc), it is challenging to study individual collapsing objects.

The majority of high-mass star formation (HMSF) takes place in the densest regions of molecular clouds. Based on observational and theoretical considerations, HMSF can be divided into several evolutionary stages \citep{Beuther2007,Zinnecker2007}: the formation of massive stars begins in infrared dark clouds (IRDCs) harboring potentially short-lived high-mass starless cores and low- to intermediate-mass protostars \citep[e.g.,][]{Pillai2006, Rathborne2006, Sanhueza2012, Zhang2015} and proceeds to form high-mass protostellar objects (HMPOs) with $M_\star > 8$\,$M_\odot$ showing gas accretion and molecular outflows \citep[e.g.,][]{Beuther2002,Motte2007}. During the hot molecular core (HMC) stage, also called the hot core stage, the central protostar(s) warm the surrounding dense envelope to temperatures higher than $ 100-300$\,K within the innermost several thousand au and rich molecular spectra at (sub)mm wavelengths are observed \citep[e.g.,][]{Belloche2013, SanchezMonge2017, Beltran2018}. At later stages, ultra-compact H{\sc ii} (UC H{\sc ii}) regions form where the protostars ionize their surrounding envelope, which is observed in strong free-free emission at cm wavelengths \citep[e.g.,][]{Garay1999,Palau2007, Qin2008,Klaassen2018}.

Numerous millimeter and submillimeter wavelength observations towards hot cores reveal a rich chemistry in nitrogen- and oxygen-bearing molecules \citep[for a review, see][]{Herbst2009} as well as sulfur-bearing species \citep{Herpin2009}. In addition, complex organic molecules (COMs) are very abundant in those environments. Following the definition by \citet{Herbst2009}, a COM consists of six or more atoms. Molecules can form in the interstellar medium (ISM), where the gas and dust are shielded from strong stellar radiation \citep[at a visual extinction of $A_\mathrm{v} \approx 200$\,mag in the dense parts of HMSFRs;][]{Tan2014}. A summary of the detected molecules in the ISM is given in \citet{McGuire2018} ranging from diatomic species such as CO up to C$_{70}$. The chemical composition of these objects can be modeled using a chemical kinetics approach \citep{Semenov2010, Cuppen2017} and an appropriate network of astrophysically relevant chemical reactions, for example, the KInetic Database for Astrochemistry \citep[KIDA,][]{Wakelam2015} and the UMIST Database for Astrochemistry \citep[UDfA,][]{McElroy2013}. The reaction data are obtained by quantum chemical calculations and ultra-high vacuum laboratory experiments \citep[for a review, see][]{Cuppen2017}.

In the early phases of star formation at low temperatures of $\sim$10\,K, atoms and small molecules (e.g., CO) produced in the gas-phase stick to the grain surface creating icy mantles \citep{Tielens1982,Hasegawa1992}. Hydrogenation of CO leads to species such as H$_2$CO and CH$_3$OH on the grains. \citet{Garrod2006} have shown that reactions between radicals on dust grains can form the most widely detected complex molecules in hot cores, such as HCOOH, CH$_3$OCHO, and CH$_3$OCH$_3$, more efficiently during a gradual warm-up phase compared to the formation in the gas-phase. Jets and molecular outflows launched from protostars and interacting with the surrounding material create shocks that temporarily increase the temperature and density \citep{Draine1993, Bachiller1996, Zhang2005, McKee2007, Anderl2013}. In such shocked regions molecular dissociation (but also reactions with activation barriers) can occur, and molecules formed on the grain mantles can desorb into the gas-phase \citep[e.g., an increased abundance of COMs can be observed;][]{Palau2017}. Silicon is mostly locked on grains, but due to powerful outflows it is sputtered off the grains and forms SiO in the gas-phase \citep{Herbst1989,Schilke1997,Gerin2013}.

The molecular richness of HMSFRs makes them ideal astrochemical laboratories. However, comprehensive chemical surveys of a large sample of high-mass star-forming cores with a high angular resolution targeting the molecular gas still need to be carried out and studied \citep[e.g.,][]{Beuther2009, Wang2014, Feng2016}. Surveys using single-dish telescopes have been made \citep[e.g.,][]{Jackson2013,Gerner2014,Gerner2015,Urquhart2019}, but their angular resolution is not sufficient to resolve individual cores and the resulting molecular abundances are beam-averaged over a large region, which may contain signals of chemically distinct cores at different evolutionary stages. Most chemical studies at high angular resolution focus on individual cores \citep[e.g.,][]{Beuther2007C, Fuente2014, Beltran2018}, whereas the influence of the environment on the chemical evolution may play an important role. For example, \citet{Feng2016}, \citet{Allen2017}, and \citet{Mills2018} observed and investigated chemical differentiation between adjacent hot cores.

In the framework of a sample of HMSFRs observed with the same setup and high spatial resolution, we carried out the NOrthern Extended Millimeter Array (NOEMA)\footnote{Upgraded version of the Plateau de Bure Interferometer (PdBI).} large program CORE at 1.37\,mm \citep{Beuther2018}. In total, 18 HMSFRs were observed with an angular resolution of $\sim$0\as4. The broad frequency bandwidth of $\sim$4\,GHz covers a large number of molecular lines and thus we are able to study the chemical properties of the objects in this sample. A detailed description of the observations is given in Sect. \ref{section:observations}. As a test case for the chemical survey, we start with a case study of AFGL\,2591 that will be complemented in the future with an analysis of the other sources in the CORE sample (Gieser et al. in prep.).

The high-mass star-forming region AFGL\,2591, located at $\alpha$(J2000) = 20:29:24.8 and $\delta$(J2000) = +40:11:19.6, has been thoroughly studied in the past. By measuring the trigonometric parallax of H$_2$O masers with the Very Long Baseline Array (VLBA), \citet{Rygl2012} determined a distance of $3.33\pm0.11$\,kpc. A multiwavelength overview of the region is shown in Fig. \ref{Fig:AFGLRegion}. In the near-infrared\footnote{Based on observations obtained at the Gemini Observatory acquired through the Gemini Observatory Archive, which is operated by the Association of Universities for Research in Astronomy, Inc., under a cooperative agreement with the NSF on behalf of the Gemini partnership: the National Science Foundation (United States), National Research Council (Canada), CONICYT (Chile), Ministerio de Ciencia, Tecnolog\'{i}a e Innovaci\'{o}n Productiva (Argentina), Minist\'{e}rio da Ci\^{e}ncia, Tecnologia e Inova\c{c}\~{a}o (Brazil), and Korea Astronomy and Space Science Institute (Republic of Korea).}, a large-scale outflow lobe can be observed in the east--west direction \citep[also Fig. 13 in][]{Hodapp2003}. At centimeter wavelengths a cluster of at least five sources is revealed with the Very Large Array \citep[VLA,][]{Campbell1984, Trinidad2003, Johnston2013}. VLA\,1 and VLA\,2 are H{\sc ii} regions \citep{Trinidad2003}. The major source in the region is the VLA\,3 hot core with a systemic velocity of $v_{\mathrm{LSR}} = -$5.5\,km\,s$^{-1}$ and a luminosity of 2$\times$10$^5$\,$L_\odot$ suggesting a protostar with $\sim$40\,$M_\odot$ \citep{Sanna2012}. VLA\,4 is a $\sim$9\,$M_\odot$ star with spectral type B2, while the faint object VLA\,5 coincides with the position of the infrared source MASS 20292393+4011105 \citep{Johnston2013}. Traced by H$_2$O maser emission, \citet{Sanna2012} and \citet{Trinidad2013} found another protostellar object, VLA\,3-N, very close to the VLA\,3 hot core ($\sim$0\as4 north).

Our analysis is focused on the hot core AFGL\,2591 VLA\,3 whose kinematic and chemical properties have been studied at multiple wavelengths in the past. Based on the kinematic properties using PdBI observations of HDO, H$_2^{18}$O, and SO$_2$, \citet{Wang2012} found a Hubble law-like expansion within the inner $2\,400$\,au, and argue that this is caused by a disk wind, a phenomenon that has been recently observed for G17.64+0.16 \citep{Maud2018}. The centimeter continuum emission was studied by \citet{vanderTakMenten2005} with the VLA. These authors modeled the emission assuming that the expansion of the H{\sc ii} region is halted by accretion flows from the molecular envelope, which reproduces the observed structure of AFGL\,2591 VLA\,3. Infrared observations with the Stratospheric Observatory for Infrared Astronomy (SOFIA) reveal molecular emission in H$_2$O and CS from the inner hot dense region \citep{Indriolo2015, Barr2018}. \citet{Jimenez2012} reported chemical segregation on scales smaller than $3\,000$\,au due to the interplay between molecular dissociation by UV radiation, ice evaporation, and high-temperature gas-phase chemistry. These authors identify three types of spatial distribution: molecules with a compact distribution at the location of the continuum peak (H$_2$S, $^{13}$CS), double-peaked distributions (HC$_3$N, OCS, SO, SO$_2$), and ring-like structures (CH$_3$OH).

We use AFGL\,2591 VLA\,3 as a case study to investigate the chemical complexity in high-mass star-forming regions based on observations at high angular resolution and using a physical-chemical model. We infer the physical structure of the source and compare it to theoretical predictions. Using the rich molecular line information, we not only quantify the chemical content, but also study the spatial distribution of the emission to address the following questions: Which molecular emission traces the compact core? Which species show extended large-scale emission? Which molecules have an asymmetric distribution? Which of those molecules could be chemically linked? What is the modeled chemical age of the hot core?

This paper is organized as follows. The CORE observations are described in Sect. 2 with a summary of the data calibration and properties of the final data products. The observational results are reported in Sect. 3, where the physical and chemical structure are analyzed. The applied physical-chemical model of the source is presented in Sect. 4. The observational and chemical modeling results are discussed in Sect. 5. In addition, the results are discussed in the context of previous studies of the target region. The main conclusions and an outlook are given in Sect. 6.
\section{Observations} \label{section:observations}

\begin{table*}[htb]
\centering
\caption{Data products of AFGL\,2591 used in this study.}
\label{tab:dataproducts}
\begin{tabular}{lcccccc}
\hline \hline
& \multicolumn{1}{c}{Beam Size} & \multicolumn{1}{c}{Position Angle} & \multicolumn{1}{c}{Frequency} & \multicolumn{2}{c}{Spectral Resolution} & \multicolumn{1}{c}{rms Noise}\\
 & $\theta_\mathrm{maj}\times\theta_\mathrm{min}$ & PA & $\nu$ & $\Delta v$ & $\Delta \nu$ & $\sigma$\\
 & ($''\times''$) & ($^\circ$) & (GHz) & (km\,s$^{-1}$) & (MHz) & \\
\hline 
\textbf{CORE data}&&&&&&\\
NOEMA continuum & 0.46$\times$0.36 & 66 & 218.873 & & & 0.59\,mJy\,beam$^{-1}$\\ 
NOEMA line data & 0.47$\times$0.36 & 66 & 217.224$-$220.773 & 3.0 & 2.2 & 1.2\,mJy\,beam$^{-1}$\,channel$^{-1}$\\ 
Merged line data & 0.47$\times$0.36 & 67 & 217.224$-$220.773 & 3.0 & 2.2 & 1.5\,mJy\,beam$^{-1}$\,channel$^{-1}$\\ 
IRAM\,30\,m LI & 11.8$\times$11.8 & 0 & 217.224$-$221.274 & 0.27 & 0.2 & 2.2\,Jy\,beam$^{-1}$\,channel$^{-1}$\\ 
IRAM\,30\,m UI & 11.2$\times$11.2 & 0 & 229.172$-$233.222 & 0.25 & 0.19 & 1.5\,Jy\,beam$^{-1}$\,channel$^{-1}$\\ 
\hline
\textbf{Archival data}&&&&&&\\
SMA SiO line data & 6.8$\times$5.6 & 28 & 216.976$-$217.214 & 1.1 & 0.8 & 30\,mJy\,beam$^{-1}$\,channel$^{-1}$\\ 
\hline 
\end{tabular} 
\end{table*}

This study is part of the NOEMA large program CORE \citep{Beuther2018}. The main goals of the project are to study of the fragmentation and disk formation processes, as well as the outflows and chemistry in high-mass star formation. A detailed description of the full survey, technical setup, scientific goals, and analysis of the continuum data is given in \citet{Beuther2018}. The observations of the sources in the 1\,mm band were carried out with NOEMA from $2014-2017$ in the A, B, and D configurations.

Using the wide-band correlator WideX, spectra from 217\,GHz to 221\,GHz with a spectral resolution of 1.95\,MHz ($\sim$2.7\,km\,s$^{-1}$ at 1\,mm) were obtained in H and V linear polarizations. In order to study the kinematics, high spectral resolution units were placed around the wavelengths of key molecular transitions \citep[see][]{Ahmadi2018, Beuther2018}. A study of the kinematics of the CORE regions, including AFGL\,2591, will be given in Ahmadi et al. (in prep.).

In order to include short-spacing information, single-dish observations with the IRAM\,30\,m telescope were also performed. The IRAM\,30\,m observations were carried out in four 4\,GHz broad spectral windows centered at $\sim$215, 219, 231, and 234\,GHz with a spectral resolution of 0.2\,MHz ($\sim$0.3\,km\,s$^{-1}$ at 219\,GHz) in H and V linear polarizations with a beam size of $\sim$12$''$. The calibration of the single-dish data and merging process of the interferometric and single-dish data are described in detail in Mottram et al. (subm.).

The following sources were observed for calibration during the AFGL\,2591 tracks: the quasars 2013+370 and 2037+511 for phase and amplitude, a strong quasar (3C454.3) for bandpass, and the star MWC349 for absolute flux calibration. The data were calibrated with the {\tt CLIC} package in {\tt gildas}\footnote{\url{http://www.iram.fr/IRAMFR/GILDAS/}}. The 1.37\,mm continuum was extracted from line-free channels in the spectral line data and then subtracted from the spectral line data. The NOEMA data (line and continuum) were deconvolved with the Clark CLEAN algorithm \citep[][]{Clark1980} as the extended emission is filtered out. The NOEMA data merged with the IRAM\,30\,m observations (hereafter merged data) were deconvolved with the SDI algorithm \citep[][]{Steer1984}, but see Mottram et al., subm., for a comparison of these two deconvolution methods. The NOEMA and merged data were smoothed to a common spectral resolution of 3\,km\,s$^{-1}$. The properties of the data products are shown in Table \ref{tab:dataproducts}.
\section{Observational analysis} \label{section:ObsAna}
\subsection{Continuum}\label{subsection:cont}
\begin{figure}[htb]
\resizebox{\hsize}{!}{\includegraphics{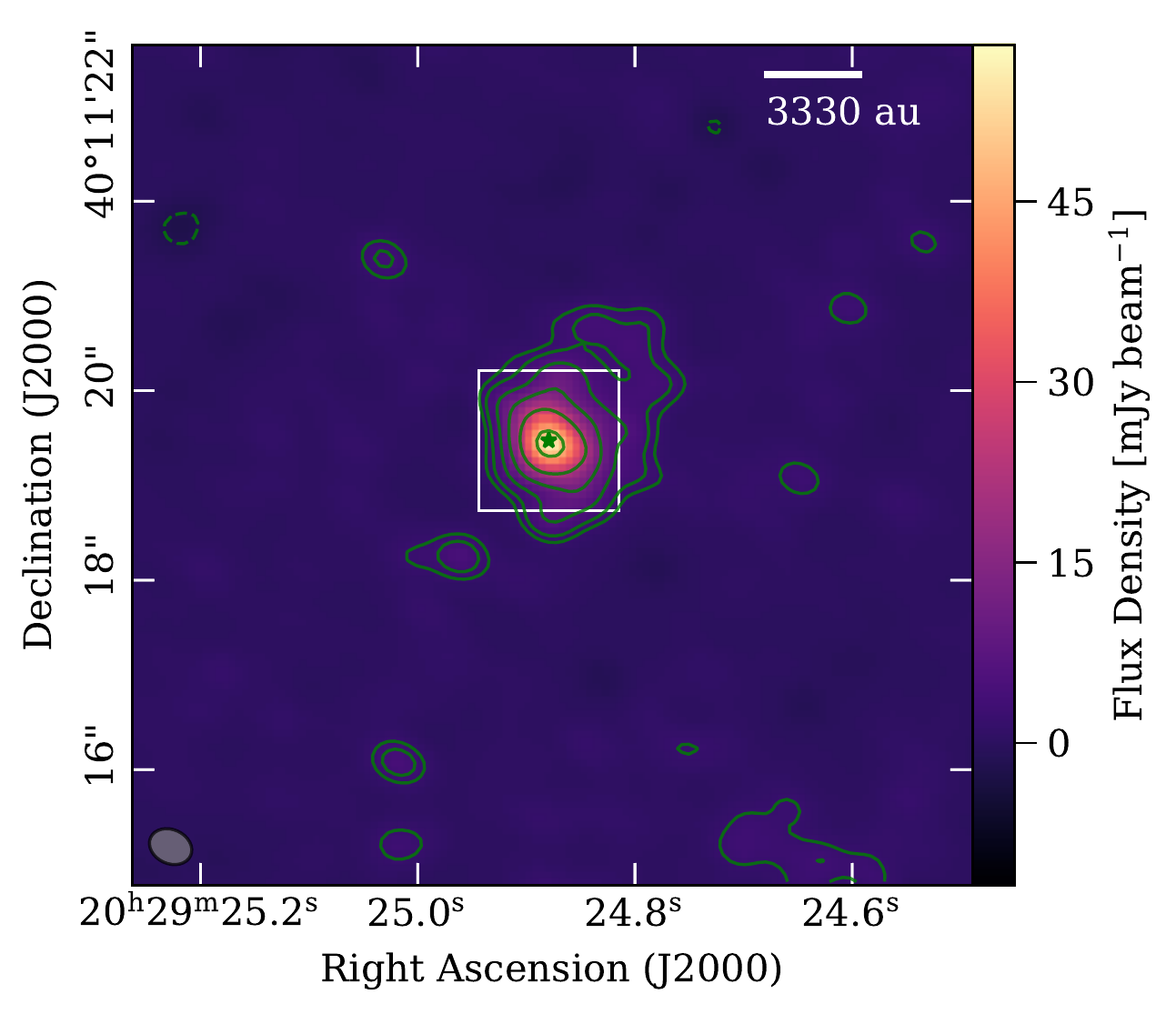}}
\caption{AFGL\,2591 continuum emission at 1.37\,mm. The green contours show the 1.37\,mm continuum emission with levels at $-$3 (dashed), 3, 5, 10, 20, 40, and 80$\sigma$ ($\sigma = 0.59$\,mJy\,beam$^{-1}$). The beam size (0\as46$\times$0\as36) is shown in the lower left corner. The green star indicates the position of the 1.37\,mm continuum peak. The white rectangle shows the region within which an average spectrum was obtained in the spectral line data for line identification (Sect.\,\ref{subsection:line_identification}).}
\label{Fig:Cont}
\end{figure}

The 1.37\,mm continuum emission of the AFGL\,2591 VLA\,3 hot core is shown in Fig. \ref{Fig:Cont}. Other VLA sources (VLA\,1, VLA\,2, and VLA\,4) are not detected within the NOEMA primary beam at a rms sensitivity of 0.59\,mJy\,beam$^{-1}$. The central emission of VLA\,3 has an approximately spherically symmetric distribution around the position of the peak flux. With an angular resolution of 0\as4 (1\,400\,au at 3.33\,kpc) we resolve the envelope, but no disk structure. Around the core there is an elongation towards the northwest, and there are several globules towards the northeast and southeast with emission $> 5 \sigma$, both of which could potentially be associated with the molecular outflow \citep[as studied by, e.g.,][]{Gueth2003}. The main contribution to the 1.37\,mm continuum emission is dust emission; based on VLA observations of H$_2$O masers in the AFGL\,2591 star-forming region, at wavelengths $\geq 1$\,cm the emission is affected by free-free emission from the ionized gas \citep{Trinidad2003}. At frequencies $\geq 43$\,GHz the emission is dominated by the dust \citep{vanderTakMenten2005}.

\subsection{Line identification}\label{subsection:line_identification}

\begin{figure*}
\centering
\includegraphics[width=17cm]{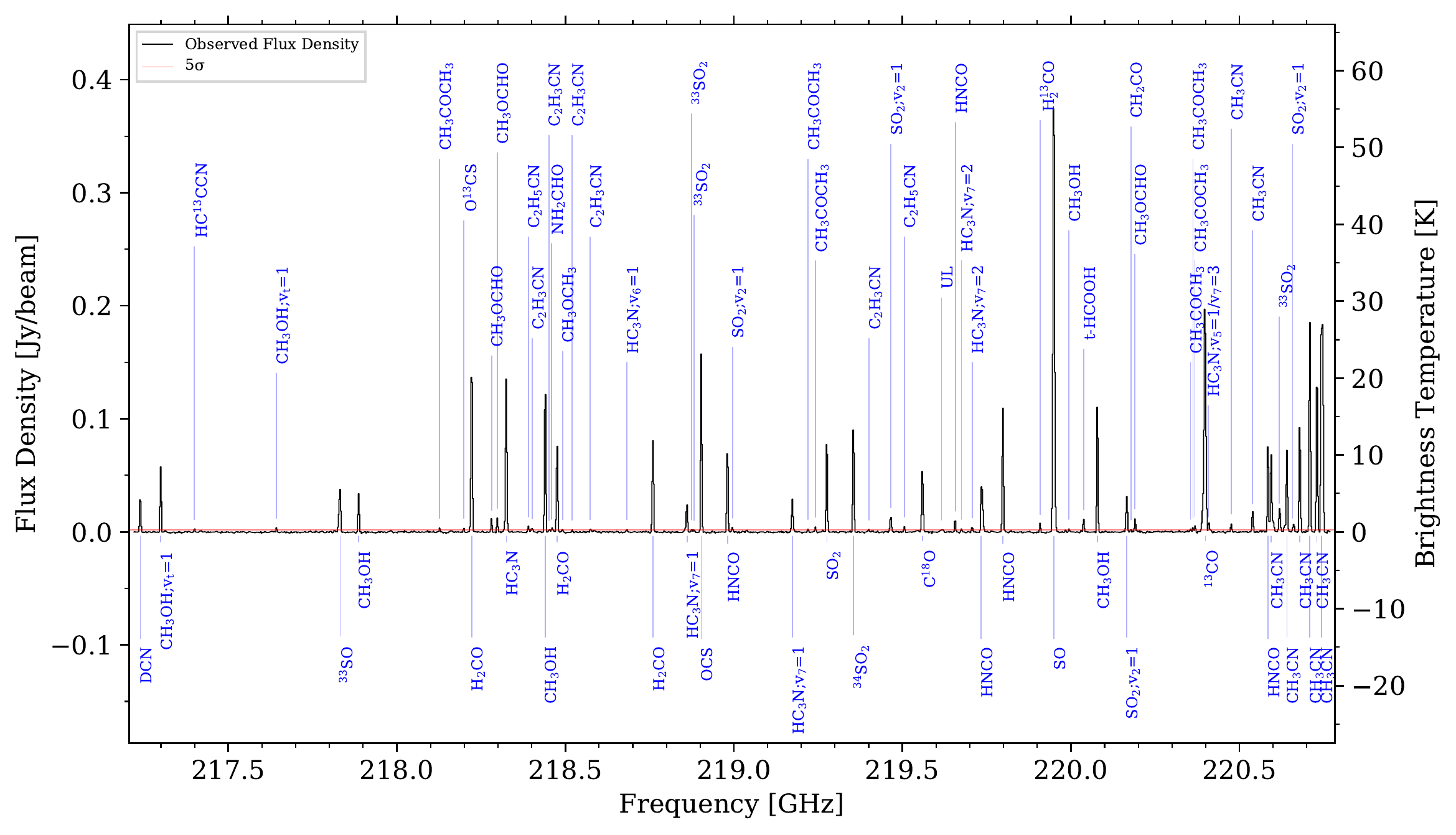}
\caption{Average spectrum (spatially averaged over 1\as6$\times$1\as6, indicated by the white rectangle in Fig. \ref{Fig:Cont}) of the AFGL\,2591 hot core labeled with molecular emission lines with $S/N > 5$ (indicated by the red horizontal line, $\sigma = 0.34$\,mJy\,beam$^{-1}$ in the average spectrum). Unidentified lines are labeled UL. A detailed table of the properties of the molecular transitions is given in Appendix \ref{sec:TransProps}.}
\label{Fig:LineID}
\end{figure*}

An average NOEMA spectrum (spatially averaged over 1\as6$\times$1\as6, indicated by the white rectangle in Fig. \ref{Fig:Cont}) around the continuum peak is used to identify emission lines detected at $S/N > 5$ ($\sigma = 0.34$\,mJy\,beam$^{-1}$ in the average spectrum). The molecular transitions are taken from the Cologne Database for Molecular Spectroscopy\footnote{\url{https://cdms.astro.uni-koeln.de/}} \citep[CDMS,][]{CDMS} and Jet Propulsion Laboratory database\footnote{\url{https://spec.jpl.nasa.gov/}} \citep[JPL,][]{JPL}. The spectrum with labeled molecular lines is shown in Fig. \ref{Fig:LineID}. The properties of a total of 67 detected lines are summarized in Table \ref{tab:tansition_props} in the Appendix. We detect many transitions from simple molecules (e.g., SO, HNCO, OCS), but also a forest of COM lines, among which CH$_3$OH and CH$_3$CN have the strongest emission lines. For some molecules (e.g., HC$_3$N and CH$_3$OH) in addition to the commonly detected rotational states, rotational transitions from vibrationally or torsionally excited states are also present. 

Multiple spectral line surveys of the AFGL\,2591 region have been carried out in the past covering partially $80 - 360$\,GHz using the James Clerk Maxwell Telescope (JCMT) and the IRAM\,30\,m telescope \citep[][]{Bisschop2007,vanderWiel2011}; and $480 - 1900$\,GHz with the Herschel space observatory \citep[][]{Ceccarelli2010, vanderWiel2013, KazmierczakBarthel2014}. \citet{Bisschop2007} and \citet{Ceccarelli2010} reported that AFGL\,2591 has a ``line-poor'' spectrum compared to other massive protostars, but with beam sizes $> 10''$ the compact emission around the hot core cannot be resolved sufficiently and is heavily beam diluted. Our 1.37\,mm spectral line data at subarcsecond resolution shows that the hot core is indeed a common hot core with a line-rich spectrum (Fig. \ref{Fig:LineID}).
 
\subsection{Physical environment of AFGL\,2591} \label{sec:physenv}
We use the high angular resolution dust continuum (NOEMA) and merged (NOEMA + IRAM\,30\,m) spectral line data of AFGL\,2591 at 1.37\,mm to analyze the physical structure of the source.

\subsubsection{Temperature structure}\label{subsec:T_profile}
The molecules H$_2$CO and CH$_3$CN can probe the kinetic temperature when multiple transitions at different energy levels are observed \citep[e.g.,][]{Zhang1998,Rodon2012}. The CH$_3$CN emission is strong around hot cores and can trace dense high-temperature gas \citep{Fuente2014}, whereas H$_2$CO is used to trace the temperature of the colder extended envelope \citep{Mangum1993}. The upper state energies $E_\mathrm{u} / k_\mathrm{B}$ of the observed transitions of H$_2$CO and CH$_3$CN are listed in Table \ref{tab:tansition_props}. 

To derive the physical parameters of the molecular gas such as column density $N$ and rotation temperature $T_\mathrm{rot}$, we fit the spectral lines with {\tt XCLASS}\footnote{\url{https://xclass.astro.uni-koeln.de/}} \citep[eXtended Casa Line Analysis Software Suite,][]{XCLASS}. The {\tt XCLASS} software models molecular lines by solving the 1D radiative transfer equation assuming local thermal equilibrium (LTE) conditions and an isothermal source. In dense sources, such as hot cores, the LTE assumption is a reasonable approximation for rotational lines; in Sect. \ref{subsec:h2gas} we estimate an average H$_2$ density of $\sim$10$^7$\,cm$^{-3}$ in the central part within a radius of $\sim$2\,500\,au. The critical density is $< 10^7$\,cm$^{-3}$ for most of the observed molecules (CH$_3$OH being the only exception; see Table \ref{tab:tansition_props}). Therefore, the population of the molecules can be described by a Boltzmann distribution with a single characteristic temperature $T_{\mathrm{ex}} = T_{\mathrm{rot}} = T_{\mathrm{kin}}$. In {\tt XCLASS}, the line profiles are assumed to be Gaussian, but optical depth effects are included. Each molecule can be described by multiple emission and absorption components. The transition properties (e.g., Einstein coefficients, state degeneracies, partition functions) are taken from an embedded SQlite3 database containing entries from CDMS and JPL using the Virtual Atomic and Molecular Data Centre \citep[VAMDC,][]{Endres2016}.

\begin{figure}[htb]
\resizebox{\hsize}{!}{\includegraphics{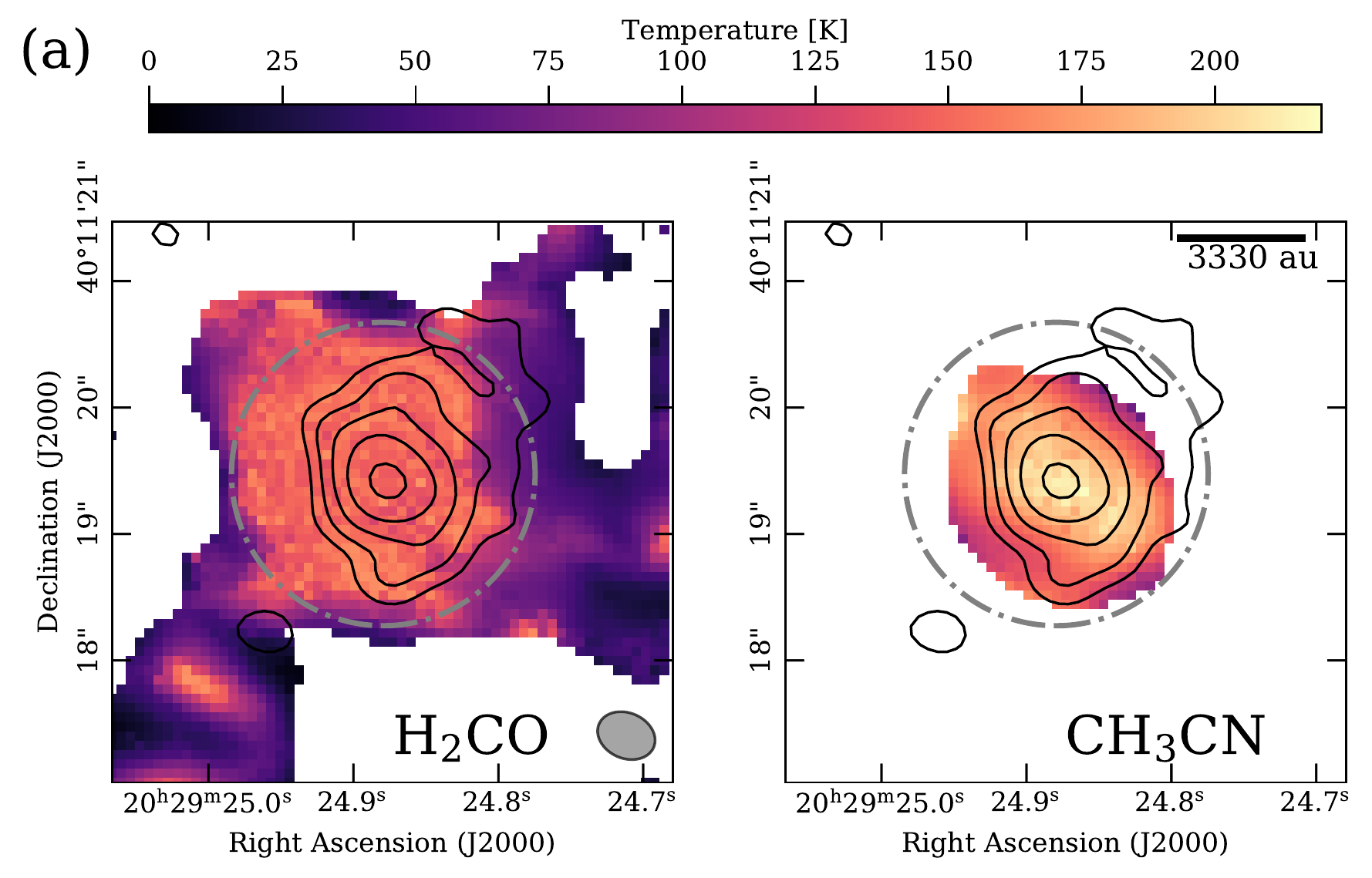}}
\resizebox{0.9\hsize}{!}{\includegraphics{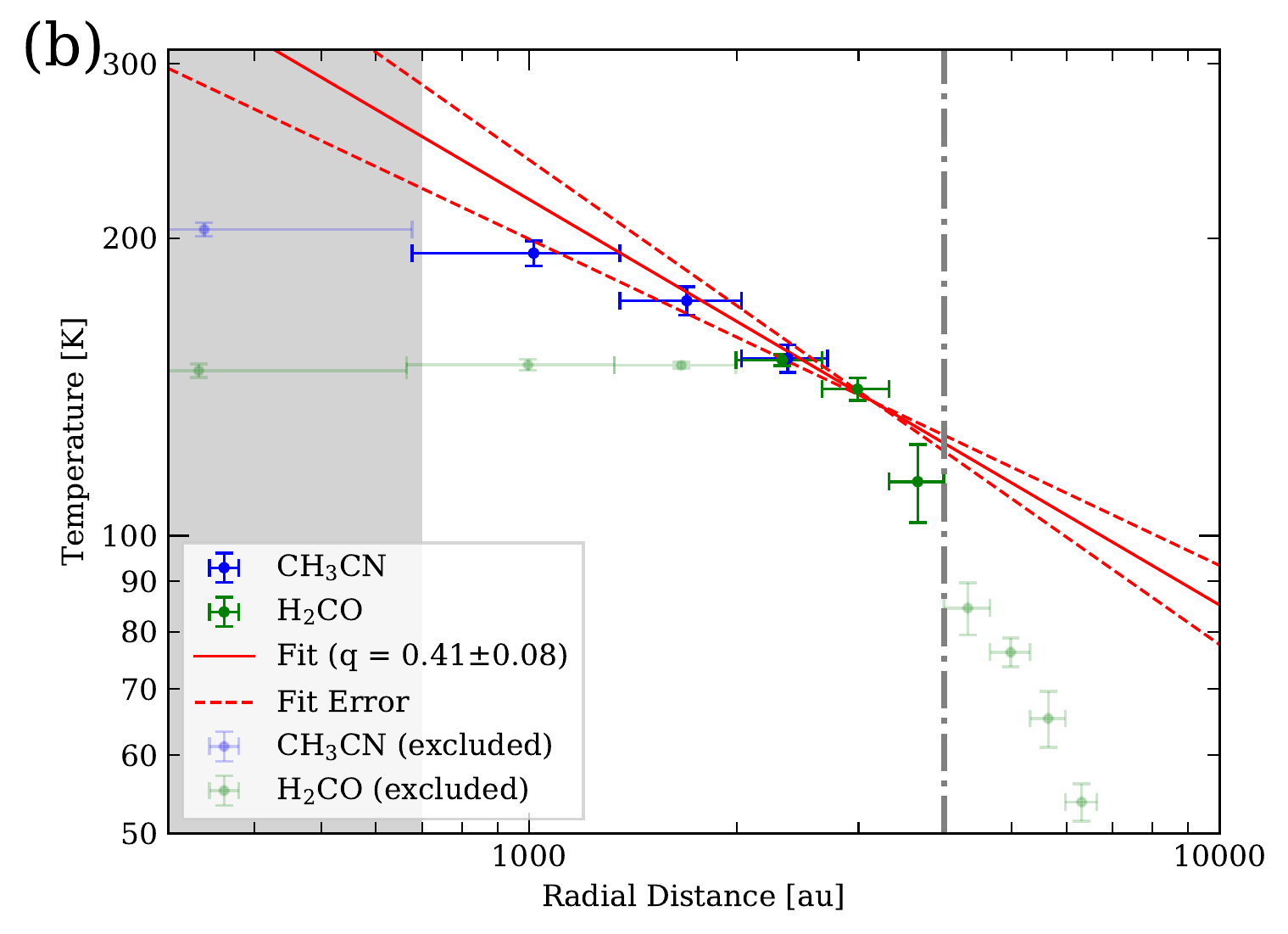}}
\caption{\textbf{(a)} H$_2$CO and CH$_3$CN temperature maps derived with {\tt XCLASS} at a threshold of 10$\sigma$. The black contours show the 1.37\,mm continuum emission with levels at 5, 10, 20, 40, and 80$\sigma$ ($\sigma = 0.59$\,mJy\,beam$^{-1}$). The gray dash-dotted circle indicates the outer radius of the radial temperature fit (bottom panel). The beam size ($0\as47\times0\as36$) is shown in the lower right corner of the H$_2$CO temperature map. \textbf{(b)} Radial temperature profiles obtained from the data in the top panel. The data are binned in steps of half a beam size ($\sim$700\,au) for H$_2$CO (green data points) and CH$_3$CN (blue data points). A power-law fit was performed on spatial scales in the range $700 - 4\,000$\,au to determine the temperature power-law index $q$. The gray dash-dotted line indicates the outer radius of the radial temperature fit. At radii smaller than 700\,au the temperature profile cannot be determined due to the finite beam size (gray shaded area). The transparent data points were excluded from the fitting. The best fit is shown by the red solid line. The error of the fit (one standard deviation) is shown by the red dashed lines.}
\label{Fig:TMap}
\end{figure}

The {\tt XCLASS} package provides new functions for CASA to model interferometric and single-dish data. By using the {\tt myXCLASSFit} function, the user can fit the model parameters (source size $\theta_\mathrm{source}$, rotation temperature $T_\mathrm{rot}$, column density $N$, linewidth $\Delta v$, and velocity offset from the systemic velocity $v_\mathrm{off}$) to observational data by using different optimization algorithms. In contrast, the {\tt myXCLASSMapFit} function can be used to fit one or more complete data cubes. After fitting the selected pixel spectra, the {\tt myXCLASSMapFit} function creates FITS images for each fitted model parameter, where each pixel corresponds to the value of the optimized parameter for this pixel. The modeling can be done simultaneously with corresponding isotopologues and rotational transitions from vibrationally excited states. The ratio with respect to the main species or ground state can be either fixed or be introduced as an additional fit parameter.

Modeling the CH$_3$CN and H$_2$CO emission lines from the merged data with one emission component in {\tt XCLASS}, we derive the temperature structure assuming $T_\mathrm{rot} = T_\mathrm{kin}$ adopting a threshold of 10$\sigma$ above the noise level ($\sigma = 1.5$\,mJy\,beam$^{-1}$ $ = 0.23$\,K). For H$_2$CO, we fit the $3_{0,3}-2_{0,3}$ and $3_{2,1}-2_{2,1}$ lines. For CH$_3$CN, we fit the $K = 4$, 5, and 6 components of the $J = 12-11$ ladder, including the weak $K = 0-3$ of CH$_3^{13}$CN transitions adopting an isotopic ratio of 60 \citep{Wilson1994}. We exclude the $K = 0-3$ components of CH$_3$CN in order to reduce errors due to the high optical depth of those transitions, which may cause self-absorption \citep[further discussed in Appendix \ref{section:linefitting} and in][]{Ahmadi2018}.

A detailed description of the optimization algorithms in {\tt XCLASS} is given in \citet{MAGIX}. It is necessary to carefully check whether the applied algorithms converge towards local minima or towards the aimed global minimum. For CH$_3$CN, an algorithm chain using the Genetic (300\,iterations) and the Levenberg–Marquardt (50\,iterations) algorithms worked best. For the H$_2$CO fitting only the Levenberg–Marquardt algorithm (50\,iterations) was used. In order to reduce the number of fit parameters from five to four, the source size $\theta_\mathrm{source}$ is fixed to 4$''$ for both molecules. In this case the beam filling factor\footnote{$\eta = \frac{\theta_\mathrm{source}^2}{\theta_\mathrm{source}^2+\theta_\mathrm{beam}^2}$} is $\sim$1 and therefore the emission is assumed to be resolved with the NOEMA beam $\theta_\mathrm{beam} \approx 0$\as42. We show in Sect. \ref{sec:moment0} that the detected molecular emission is resolved in our data.

The derived H$_2$CO and CH$_3$CN temperature maps are shown in Fig. \ref{Fig:TMap}(a). The temperature distribution of CH$_3$CN is approximately spherically symmetric and is centered around the 1.37\,mm continuum peak tracing the dense inner core with temperatures up to $\sim$200\,K, while H$_2$CO shows more extended emission tracing the colder envelope with lower temperatures ranging from 30\,K to 150\,K. In the case of H$_2$CO, we restrict the analysis to the central area of $\sim$3$''\times$3$''$ as the outer edges of the H$_2$CO maps are sensitivity limited. For CH$_3$CN this is not an issue as the emission is less extended. A discussion of the uncertainties of the fit parameters is given in Sect. \ref{sec:XCLASSFittPeak}. In the central region, the H$_2$CO temperature distribution reaches a noisy plateau at temperatures $> 100$\,K. This is due to the fact that the observed H$_2$CO lines cannot trace the hottest gas \citep[as seen in Fig. 7 in][]{Rodon2012}. As the rotation temperature is determined from the line ratios, its estimate would be more accurate if more than three transitions at different energy levels could have been observed. With only two lines at different upper energy levels (21\,K and 68\,K), the fitting parameters $N$ and $T_\mathrm{rot}$ are degenerate.

Assuming spherical symmetry, we derive the radial temperature profile of both maps with the position of the 1.37\,mm continuum peak emission as the center at $\alpha$(J2000) = 20:29:24.88 and $\delta$(J2000) = +40:11:19.47. The radial temperature profiles are shown in Fig. \ref{Fig:TMap}(b) and there is a consistent overlap between the derived temperatures of both molecules at $\sim$2\,000\,au. The inner part below 700\,au is spatially unresolved in our observations. The slope of the H$_2$CO temperature profile gets steeper in the outer region at radii $> 4\,000$\,au. These data points are not included in the fit, as a large fraction of the H$_2$CO line fluxes are below the adopted threshold of 10$\sigma$ with a clear deviation from spherical symmetry as seen in Fig. \ref{Fig:TMap}(a). By combining and fitting (minimum $\chi ^2$ method) the CH$_3$CN and H$_2$CO temperature profiles, we derive the radial temperature profile from radii ranging from 700\,au to 4\,000\,au: 
\begin{equation}\label{eq:obsT}
T(r) = (255\pm165)\,\mathrm{K} \times \bigg( \frac{r}{691\,\mathrm{au}} \bigg) ^{-q}~\mathrm{, with}~q = 0.41\pm0.08.
\end{equation}
 
\citet{Palau2014} determined the temperature structure using a completely independent method fitting simultaneously the spectral energy distribution (SED) and radial intensity profiles. These authors find a value of $q = 0.40\pm0.01$ and a temperature of $250\pm20$\,K at 1\,000\,au for AFGL\,2591 (their Tables 1 and 2). In a recent infrared study by \citet{Barr2018}, hot abundant CS is detected with a temperature of $714\pm59$\,K from a region $< 130$\,au. This is consistent with the temperature profile from our study, where this temperature is reached at a radius of 56\,au.

\subsubsection{Density structure}\label{subsec:densityprofile}
The peak flux density of the 1.37\,mm continuum is 55.47\,mJy\,beam$^{-1}$ corresponding to a brightness temperature of 8.5\,K (Fig. \ref{Fig:Cont}), whereas the gas has temperatures $> 200$\,K (Fig. \ref{Fig:TMap}). We can therefore assume that the 1.37\,mm continuum emission is optically thin and can thus be used as a tracer of the H$_2$ column density which we analyze in the next section. However, because the continuum data are missing short-spacings, the images cannot be used to derive reliable intensity profiles. To overcome this problem, the intensity profiles can be fit in the Fourier-plane \citep[e.g.,][]{Beuther2007B}. The density profile can be determined using the previously derived temperature power-law index $q$ and the measured continuum intensity distribution derived from the visibilities $V(s)$ as a function of the $uv$ distance $s$. Following the theoretical work of \citet{Adams1991} and \citet{Looney2003}, the density power-law index $p$ can be derived assuming spherical symmetry:
\begin{equation}
V(s) \propto s^{p + q - 3} \propto s^{\alpha}.
\label{Eq:density_powerlaw_index}
\end{equation}

\begin{figure}[htb]
\resizebox{\hsize}{!}{\includegraphics{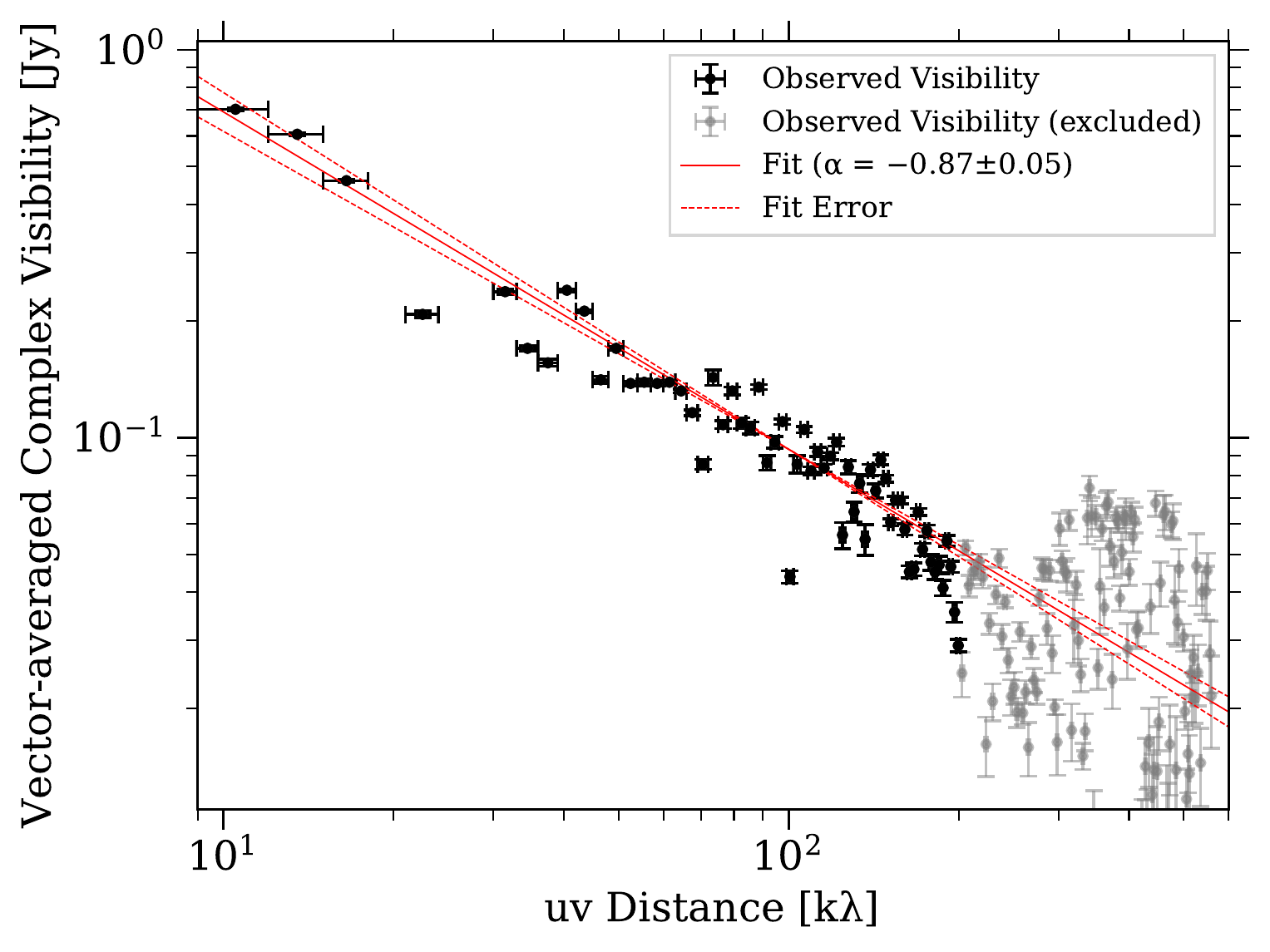}}
\caption{Vector-averaged complex visibilities $V$ of the 1.37\,mm continuum as a function of $uv$ distance. A power-law fit (red solid line) was performed in order to derive the power-law index $\alpha$. The gray data points were excluded from the fit. The error on the fit (one standard deviation) is shown by the red dashed lines.}
\label{Fig:VisibilityFit}
\end{figure}

The vector-averaged complex visibilities of the 1.37\,mm continuum emission, computed using the {\tt miriad} software \citep{miriad}, are plotted against the $uv$ distance in Fig. \ref{Fig:VisibilityFit}. The visibilities were shifted such that the position of the 1.37\,mm continuum peak is at the phase center at $\alpha$(J2000) = 20:29:24.88 and $\delta$(J2000) = +40:11:19.47. With a power-law fit (minimum $\chi ^2$ method), an index of $\alpha = -0.87\pm0.05$ is obtained on a physical scale from $\sim$3\,000$-$67\,000\,au. Visibilities with a uv distance $s > 200$\,k$\lambda$ are excluded from the fit due to an increase in the noise level. Combined with the previous determination of the temperature index $q$ (Sect. \ref{subsec:T_profile}), the density index $p$ can be calculated according to Eq. \ref{Eq:density_powerlaw_index} as
\begin{equation}
p = \alpha - q + 3.
\end{equation}
We obtain a density power-law index of $p = 1.7\pm0.1$. This result agrees with \citet{Palau2014}, where these authors find $p = 1.80\pm0.03$ for AFGL\,2591. Observations of HMSFRs find similar values for the density power-law index \citep[e.g.,][]{vanderTak2000, Beuther2002, Mueller2002, Hatchell2003, Beuther2007B, Zhang2009} and a detailed discussion is given in Sect. \ref{sec:disphysstruc}.
\subsubsection{H$_2$ column density and gas mass}\label{subsec:h2gas}
Optically thin dust emission dominates in the millimeter and submillimeter regime as discussed in the previous section. As noted in Sect. \ref{subsection:cont}, the contribution from the ionized gas around AFGL\,2591 can be neglected at 1.37\,mm. Assuming that the gas and dust are thermally coupled, the molecular hydrogen column density $N(\mathrm{H}_2)$ can be determined \citep{Hildebrand1983}:
\begin{equation}
N(\mathrm{H}_2) = \frac{S_{\nu} \eta }{\mu m_{\mathrm{H}} \Omega \kappa_{\nu} B_{\nu}(T)}.
\label{Eq:N}
\end{equation}
Here $S_{\nu}$ is the flux density, $m_{\mathrm{H}}$ the mass of a hydrogen atom, $\Omega$ the beam solid angle, and $B_{\nu}$($T$) the Planck function. Throughout the paper, we assume a gas-to-dust mass ratio of $\eta = 150$ \citep{Draine2011}, a mean molecular weight of $\mu = 2.8$, and a dust opacity of $\kappa_{\nu} = 0.9$\,cm$^2$\,g$^{-1}$ \citep[for dust grains with a thin icy mantle at 1.3\,mm at a gas density of 10$^6$\,cm$^{-3}$; Table 1 in][]{Ossenkopf1994}. In each pixel, $T$ is taken from the CH$_3$CN and H$_2$CO temperature maps shown in Fig. \ref{Fig:TMap}(a) with $T = T$(CH$_3$CN) if $T$(CH$_3$CN) $> T$(H$_2$CO) and $T = T$(H$_2$CO) elsewhere. Due to the temperature and density gradients, this is a rough approximation as CH$_3$CN traces the inner dense region, while H$_2$CO traces the colder envelope.

\begin{figure}[htb]
\resizebox{\hsize}{!}{\includegraphics{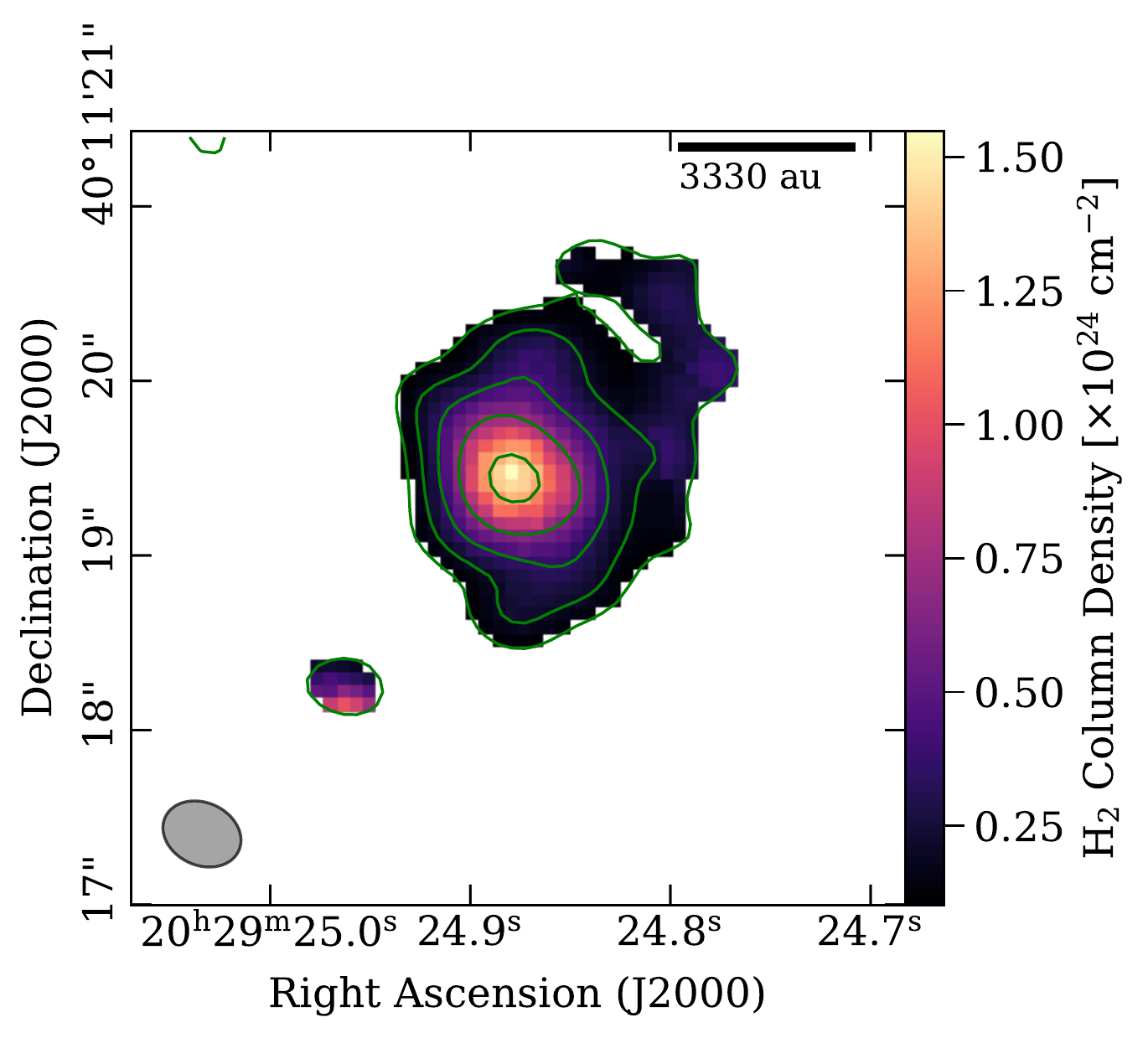}}
\caption{Column density distribution of H$_2$ derived from the 1.37\,mm continuum emission at a $5\sigma$ threshold. The green contours show the 1.37\,mm continuum emission with levels at 5, 10, 20, 40, and 80$\sigma$ ($\sigma = 0.59$\,mJy\,beam$^{-1}$). The beam size ($0\as46\times0\as36$) is shown in the lower left corner.}
\label{Fig:H2ColDens}
\end{figure}

The total gas mass can be derived as
\begin{equation}
M_{\mathrm{gas}} = \frac{F_{\nu} d^2 \eta}{\kappa_{\nu} B_{\nu}(T)},
\label{Eq:Mass}
\end{equation}
where $F_{\nu}$ is the integrated flux density and $d$ denotes the distance \citep[$3.33\pm0.11$\,kpc,][]{Rygl2012}.

The resulting H$_2$ column density distribution based on Eq. \ref{Eq:N} is shown in Fig. \ref{Fig:H2ColDens}. The H$_2$ column density reaches 1.5$\times$10$^{24}$\,cm$^{-2}$ in the center. Assuming a spherical source with a diameter of $\sim$5\,000\,au (1.5$''$) along the line of sight as traced by our data in Fig. \ref{Fig:Cont}, the average H$_2$ volume density in the center is 2$\times$10$^7$\,cm$^{-3}$. 

Using Eq. \ref{Eq:Mass}, the integrated total gas mass of the hot core core is $6.9\pm0.5$\,$M_\odot$ for emission $> 5\sigma$ within the innermost 10\,000\,au, assuming that the error budget is dominated by the uncertainties of the distance and flux density. A missing flux of 84\,\% due to spatial filtering \citep{Beuther2018}, means that this should be taken as a lower limit. With continuum SED modeling of submillimeter JCMT SCUBA observations, \citet{vanderTak2013} estimate a gas mass of 373\,$M_\odot$ within a radius of $7.1\times10^4$\,au. Based on SCUBA observations with a 22\as9 beam at 850\,$\mu$m \citep{DiFrancesco2008}, the gas mass of the complete large-scale clump is estimated to be 638\,$M_\odot$ (using the same values for $\kappa_{\nu}$ and $\eta$, and an average temperature of the region of $T = 69$\,K derived from the IRAM\,30\,m data) within an effective radius of 77\as6 corresponding to 1.3\,pc \citep[Table 1 in][]{Beuther2018}. This suggests that the AFGL\,2591 hot core is embedded in a large-scale envelope providing a large gas reservoir.

\subsubsection{Outflow(s)}\label{subsec:outflow}

\begin{figure*}
\centering
\includegraphics[width=17cm]{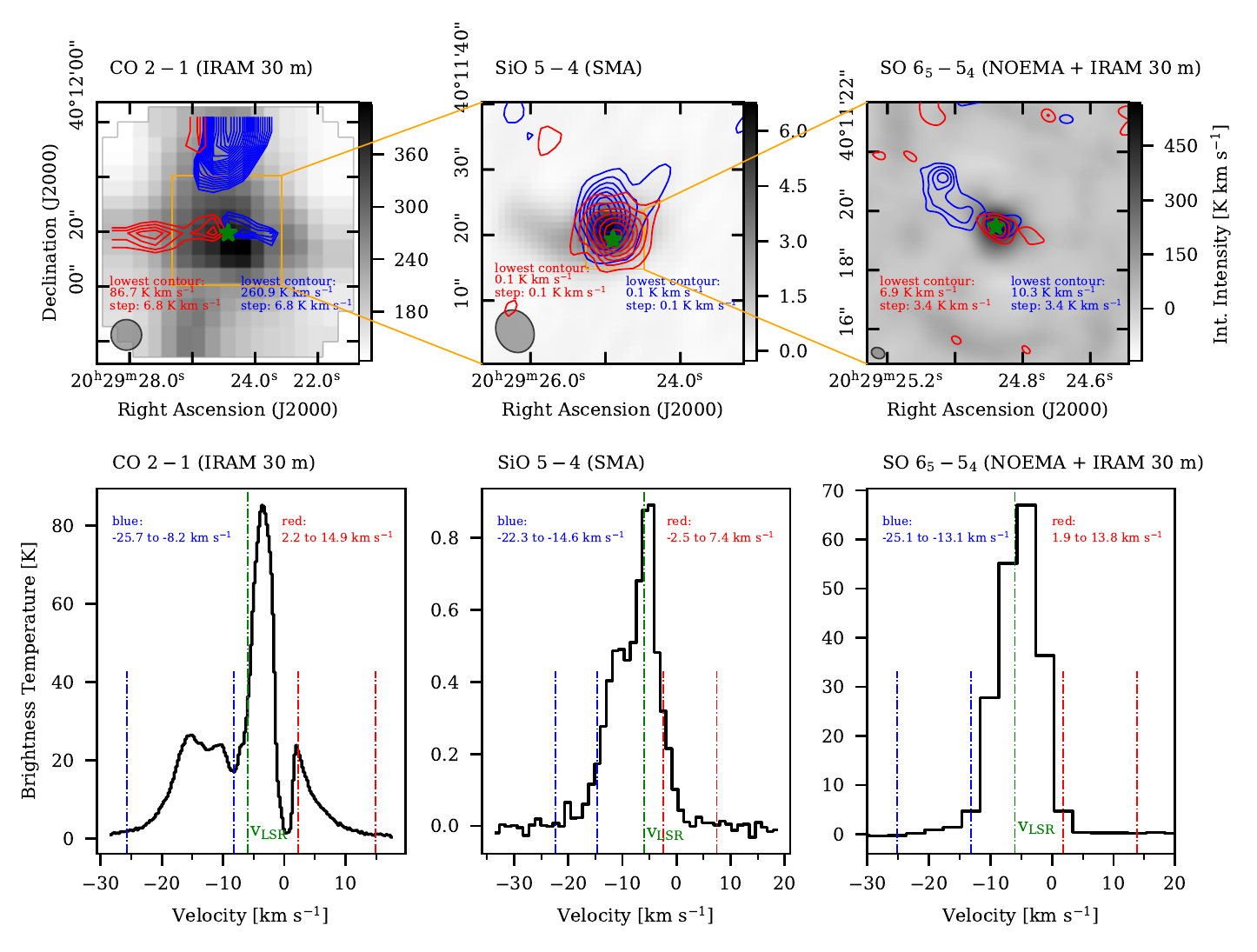}
\caption{Outflow structure of AFGL\,2591 at different spatial scales traced by IRAM\,30\,m observations of CO (left panels), SMA observations of SiO (central panels), and NOEMA + IRAM\,30\,m observations of SO (right panels). \textbf{Upper panels:} The background gray scale shows the integrated intensity of the central line emission. The blue and red contours show the blue- and redshifted line emission. The beam size is given in the lower left corner in each panel. The green star marks the position of the 1.37\,mm continuum peak. The orange box indicates the change of field of view. \textbf{Lower panels:} Spectra of the outflow tracers extracted from the position of the 1.37\,mm continuum peak (green star). The systemic velocity of $-$5.5\,km\,s$^{-1}$ is indicated by the green dashed line. The blue and red dashed lines show the integration intervals for the blue- and redshifted line emission. The integration interval of the central line emission is set between the two ranges.}
\label{Fig:Outflow}
\end{figure*}

Infrared observations reveal a large-scale east--west outflow from the VLA\,3 hot core as shown in Fig. \ref{Fig:AFGLRegion}. In this section, we investigate how this compares to the outflow morphology traced by our high angular resolution observations. The outflow structure of AFGL\,2591, as traced on various spatial scales in CO, SiO, and SO emission, is presented in Fig. \ref{Fig:Outflow}. The properties of the different data products are summarized in Table \ref{tab:dataproducts}.

The left panel of Fig. \ref{Fig:Outflow} shows the CO $2-1$ emission obtained by the IRAM\,30\,m telescope tracing scales down to $\sim$37\,000\,au. There is a clear east--west(redshifted--blueshifted) bipolar outflow with the position of the 1.37\,mm continuum peak in the center. The emission 30$''$ north originates from the large-scale filament in which the AFGL\,2591 star-forming region is embedded. The observed outflow structure of AFGL\,2591 VLA\,3 is consistent with subarcsecond resolution observations of the same line by \citet{Jimenez2012}, infrared observations (shown in Fig. \ref{Fig:AFGLRegion}), CO $3-2$ observations by \citet{Hasegawa1995}, and H$_2$O maser observations by \citet{Sanna2012}.

Another outflow tracer is SiO, which is produced by shocks \citep{Schilke1997}. No SiO line is present in the WideX frequency setup of the CORE observations with NOEMA. Archival Submillimeter Array\footnote{The Submillimeter Array is a joint project between the Smithsonian Astrophysical Observatory and the Academia Sinica Institute of Astronomy and Astrophysics and is funded by the Smithsonian Institution and the Academia Sinica.} (SMA) observations of the SiO $5-4$ line at 217.105\,GHz (project number 2010B-S108) are presented in the central panel in Fig. \ref{Fig:Outflow} tracing scales down to $\sim$21\,000\,au (Jim{\'e}nez-Serra, priv. comm.). The redshifted emission peaks at the position of the 1.37\,mm continuum peak, while the blueshifted part peaks towards the north having an elongated structure towards the northwest direction. The integrated intensity around the central line shown by the gray scale reveals an extended structure towards the east. Within the SiO data at intermediate angular resolution, no clear outflow direction can be identified. 

The right panel of Fig. \ref{Fig:Outflow} shows the outflow structure traced by the SO $6_5-5_4$ line obtained by the merged (NOEMA + IRAM\,30\,m) data tracing scales down to $\sim$1\,400\,au. The redshifted outflow emission peaks towards the 1.37\,mm continuum peak is similar to what is seen in SiO with a faint elongation towards the west. The blueshifted outflow part shows a clear direction towards the east in contrast to the large-scale outflow observed in CO where the blueshifted outflow points to the opposite direction.

The outflow has different morphologies using several outflow tracers at different spatial scales as shown in Fig. \ref{Fig:Outflow}. While the large-scale east--west(redshifted--blueshifted) outflow traced by CO agrees with previous observations, at subarcsecond resolution the SO line might trace an east--west(blueshifted--redshifted) outflow. A similar outflow morphology is found for the intermediate-mass hot core IRAS 22198+6336 with two bipolar outflows: one of the outflows is traced by SiO emission, while the other outflow is traced by HCO$^{+}$ emission \citep{SanchezMonge2011}. While we do not resolve individual fragmented collapsing objects in the central region, different outflow directions traced by different molecules may indicate that AFGL\,2591 VLA\,3 hosts multiple chemically distinct young stellar objects (YSOs) with outflows. The outflow of the VLA\,3-N object is traced by bow-shock structures of H$_2$O maser emission and has a north--south direction \citep{Sanna2012,Trinidad2013}. This source could be a good candidate to explain the anomalous direction of the SiO emission compared to the direction of the more evolved CO outflow. The inferred timescales for the H$_2$O maser bow-shocks are only 14\,years, as inferred by \citet{Sanna2012}, which is consistent with the idea that this second outflow might be very young and hence we see it only in SiO. The emission might also be caused by a disk wind, as studied by \citet{Wang2012}, for the AFGL\,2591 hot core. Observations of outflow tracers at a higher angular resolution are required in order to investigate the outflow morphology in greater detail, which will be possible in the future by NOEMA.

\subsection{Molecular content towards the continuum peak}\label{sec:XCLASSFittPeak}

\begin{table}[htb]
\caption{{\tt XCLASS} fitting results towards the 1.37\,mm continuum peak position.}
\label{tab:ColDensT}
\centering
{\renewcommand{\arraystretch}{1.2}
\begin{tabular}{lcc}
\hline \hline
Molecule & Rotation Temperature & Column Density \\
 & $T_{\mathrm{rot}}$ & $N$ \\
 &(K)&(cm$^{-2}$)\\
\hline
$^{13}$CO & $148\pm43$ & $8.9(17)\pm2.7(17)$\\ 
C$^{18}$O & $42\pm28$ & $8.0(16)\pm4.9(16)$\\ 
SO & $80\pm10$ & $1.9(16)\pm7.1(17)$\\ 
$^{33}$SO & $59\pm40$ & $9.0(14)\pm5.1(14)$\\ 
DCN & $82\pm54$ & $3.2(13)\pm3.6(12)$\\ 
OCS & $109\pm33$ & $1.7(16)\pm2.1(16)$\\ 
SO$_{2}$ & $169\pm5$ & $3.8(17)\pm8.0(16)$\\ 
$^{33}$SO$_{2}$ & $84\pm43$ & $1.8(15)\pm3.2(14)$\\ 
SO$_{2}$;$v_2$=1 & $172\pm12$ & $6.6(17)\pm1.5(17)$\\ 
H$_{2}$CO & $167\pm20$ & $1.1(16)\pm1.6(15)$\\ 
HNCO & $149\pm1$ & $7.0(15)\pm1.5(15)$\\ 
HC$_{3}$N & $212\pm83$ & $2.1(15)\pm5.4(14)$\\ 
HC$_3$N;$v_7$=1 & $137\pm19$ & $2.8(15)\pm7.7(14)$\\ 
t-HCOOH & $71\pm12$ & $1.5(15)\pm4.6(14)$\\ 
CH$_3$CN & $199\pm1$ & $4.5(15)\pm9.3(14)$\\ 
CH$_3$OH & $162\pm2$ & $9.2(16)\pm1.9(16)$\\ 
CH$_3$OH;$v_t$=1 & $110\pm4$ & $2.9(17)\pm4.4(16)$\\ 
CH$_3$OCHO & $82\pm18$ & $4.9(15)\pm1.5(15)$\\ 
C$_2$H$_5$CN & $218\pm2$ & $7.5(14)\pm1.5(14)$\\ 
\hline
\end{tabular}
}
\tablefoot{a(b) = a$\times$10$^{\mathrm{b}}$.
}
\end{table}

The spectrum towards the position of the 1.37\,mm continuum peak, extracted from the merged (NOEMA + IRAM\,30\,m) spectral line data, is used to quantify the line emission properties of all detected molecules: rotation temperature $T_\mathrm{rot}$, column density $N$, linewidth $\Delta v$, and velocity offset $v_\mathrm{off}$ from the systemic velocity. A detailed description of the {\tt XCLASS} fitting procedure and error estimation using {\tt myXCLASSFit} is given in Appendix \ref{section:linefitting}. In contrast to the core-averaged spectrum shown in Fig. \ref{Fig:LineID}, the $S/N$ of the HC$_3$N;$v_5$=1/$v_7$=3, CH$_2$CO, NH$_2$CHO, CH$_3$COCH$_3$, and C$_2$H$_3$CN emission is too low to fit the lines with {\tt XCLASS} in the spectrum extracted at the position of the 1.37\,mm continuum peak. The HC$^{13}$CCN and $^{13}$CH$_3$OH lines are blended and are therefore excluded from the fitting as well. We show in Sect. \ref{sec:moment0} that the molecular emission is spatially resolved, hence we adopt a beam filling factor $\eta$ of 1 by fixing the source size $\theta_\mathrm{source}$ to 4$''$. This also reduces the fit parameter set from five to four. With {\tt XCLASS}, it is also possible to model isotopologues simultaneously in a single fit. Then the rotation temperature, linewidth, and velocity offset are modeled together, and the column density is calculated according to the assumed isotopic ratios. The following isotopologues were fitted simultaneously with their main species: O$^{13}$CS, $^{34}$SO$_2$, H$_2^{13}$CO, CH$_3^{13}$CN. We apply fixed isotopic ratios taken from \citet{Wilson1994}: $^{12}$C/$^{13}$C $= 60$ and $^{32}$S/$^{34}$S $= 22$. As these authors do not discuss the $^{32}$S/$^{33}$S isotopic ratio, the $^{33}$S isotopologues are fitted individually.

The results for the rotation temperature and column densities are given in Table \ref{tab:ColDensT}. The spectral resolution of 3\,km\,s$^{-1}$ of the merged data is too low to study in detail the kinematic properties of the emission lines, but is sufficient to determine the rotation temperature and column density. The mean linewidth $\Delta v$ of the detected molecules is $5.2\pm0.6$\,km\,s$^{-1}$ and the mean velocity offset $v_\mathrm{off}$ from the systemic velocity ($v_{\mathrm{LSR}} = -$5.5\,km\,s$^{-1}$) is $0.0\pm0.3$\,km\,s$^{-1}$. A comparison between the observed and modeled best-fit spectrum is shown in Fig. \ref{Fig:XCLASSFit} in the Appendix. If only one strong transition is present in the spectrum, there is a large degeneracy between the column density and rotation temperature, thus the uncertainties of the fit parameters are high (e.g., for $^{13}$CO). Due to the large optical depth of the SO line (bottom panel in Fig. \ref{Fig:XCLASSFit}), the derived column density can only be taken as a lower limit. However, if many transitions with different energy levels of a molecule are detected, the modeling with {\tt XCLASS} is more reliable (e.g., for CH$_3$CN and HNCO). The median temperature uncertainty is $\sim$14\,\% and the median column density uncertainty is $\sim$22\,\%. We calculate the average $^{32}$S/$^{33}$S isotopic ratio using SO$_2$ and $^{33}$SO$_2$ to be $211\pm58$. Using CS emission, \citet{Chin1996} found that for an ensemble of star-forming regions $^{32}$S/$^{33}$S $ = 153\pm40$, which is in agreement with our results. The $^{32}$S/$^{33}$S isotopic ratio in the solar system has a lower value of 127 \citep{Anders1989}.
\subsection{Spatial molecular distribution}\label{sec:moment0}

\begin{figure*}
\centering
\includegraphics[width=17.4cm]{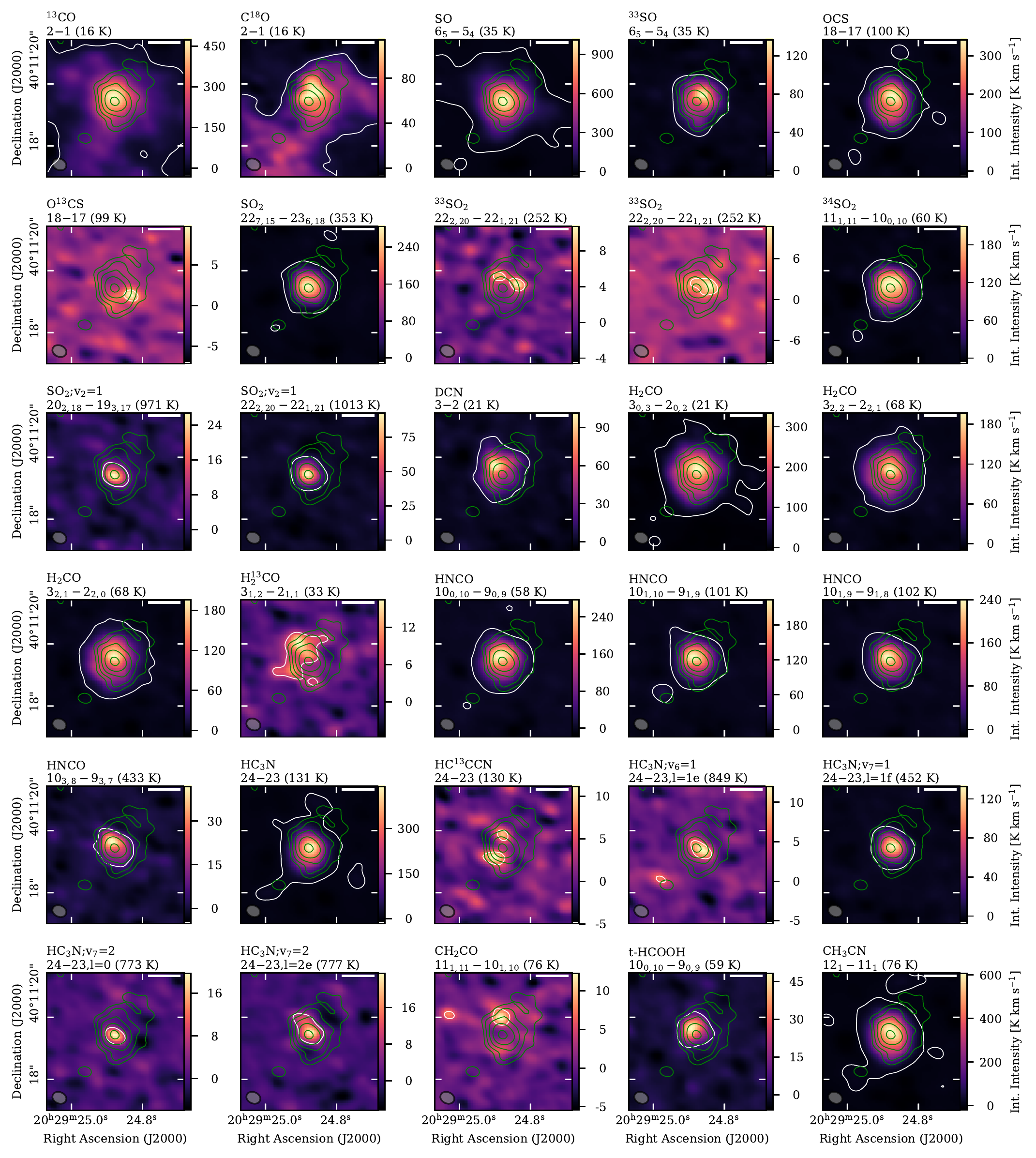}
\caption{Integrated intensity maps of detected molecular emission lines. In each panel the molecule and transition are noted (see Table \ref{tab:tansition_props}). The color map shows the integrated intensity of the line emission. The white contours indicate the 5$\sigma$ level of the integrated intensity ($\sigma = 1.2$\,K\,km\,s$^{-1}$). The green contours show the 1.37\,mm continuum emission with levels at 5, 10, 20, 40, and 80$\sigma$ ($\sigma = 0.59$\,mJy\,beam$^{-1}$). The beam size of the spectral line data ($0\as47\times0\as36$) is shown in the lower left corner. The white bar in the top right corner indicates a spatial scale of 3\,330\,au (1$''$).}
\label{Fig:Moment0_1}
\end{figure*}

\begin{figure*}
\centering
\includegraphics[width=17.4cm]{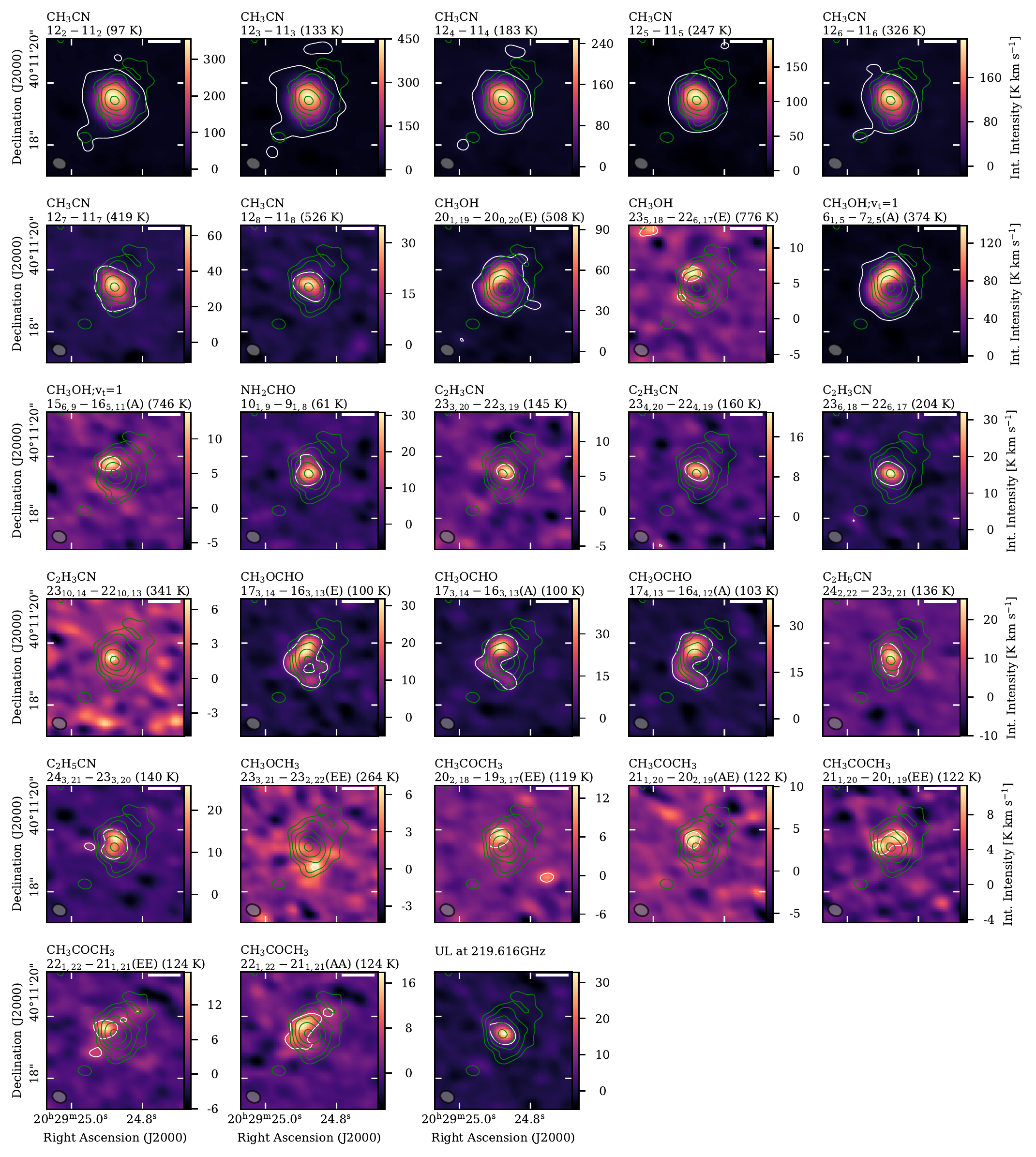}
\caption{Integrated intensity maps of detected molecular emission lines. In each panel the molecule and transition are noted (see Table \ref{tab:tansition_props}). The color map shows the integrated intensity of the line emission. The white contours indicate the 5$\sigma$ level of the integrated intensity ($\sigma = 1.2$\,K\,km\,s$^{-1}$). The green contours show the 1.37\,mm continuum emission with levels at 5, 10, 20, 40, and 80$\sigma$ ($\sigma = 0.59$\,mJy\,beam$^{-1}$). The beam size of the spectral line data ($0\as47\times0\as36$) is shown in the lower left corner. The white bar in the top right corner indicates a spatial scale of 3\,330\,au (1$''$).}
\label{Fig:Moment0_2}
\end{figure*}

\begin{table}[htb]
\caption{Positions of the peak integrated intensity taken from the data shown in Fig. \ref{Fig:Moment0_1} and Fig. \ref{Fig:Moment0_2} and classification of the morphology.}
\label{tab:Moment0_Pos}
\begin{scriptsize}
\centering
\setlength{\tabcolsep}{5pt}
\begin{tabular}{lccccl}
\hline \hline
\multicolumn{1}{l}{Molecule} & \multicolumn{1}{c}{Transition} & \multicolumn{2}{c}{Spatial Offset$^{a}$} & \multicolumn{1}{c}{Peak Int. Intensity} & \multicolumn{1}{c}{Type$^{b}$}\\
 & & $\Delta\alpha$ & $\Delta\delta$ & $\int S$ d$v$ & \\
 & & ($''$) & ($''$) & (K\,km\,s$^{-1}$)&\\
\hline
$^{13}$CO & 2$-$1 & 0.0 & 0.07 & 473.7 & I P\\ 
C$^{18}$O & 2$-$1 & -0.07 & 0.15 & 114.2 & I N\\ 
SO & 6$_{5}-$5$_{4}$ & -0.07 & -0.07 & 1013.1 & I P\\ 
$^{33}$SO & 6$_{5}-$5$_{4}$ & -0.22 & 0.15 & 136.9 & I NW\\ 
OCS & 18$-$17 & -0.07 & 0.07 & 341.5 & I P\\ 
O$^{13}$CS & 18$-$17 & -0.52 & -0.3 & 9.7 & I SW\\ 
SO$_2$ & 22$_{7,15}-$23$_{6,18}$ & 0.0 & 0.0 & 284.7 & I P\\ 
$^{33}$SO$_2$ & 22$_{2,20}-$22$_{1,21}$ & -0.52 & 0.07 & 10.7 & II W\\ 
$^{33}$SO$_2$ & 22$_{2,20}-$22$_{1,21}$ & -0.44 & 0.0 & 10.7 & I W\\ 
$^{34}$SO$_2$ & 11$_{1,11}-$10$_{0,10}$ & 0.0 & 0.07 & 209.2 & I P\\ 
SO$_2$;$v_2$=1 & 20$_{2,18}-$19$_{3,17}$ & 0.0 & -0.07 & 26.7 & I P\\ 
SO$_2$;$v_2$=1 & 22$_{2,20}-$22$_{1,21}$ & 0.0 & -0.07 & 92.5 & I P\\ 
DCN & 3$-$2 & 0.22 & 0.22 & 101.3 & I NE\\ 
H$_2$CO & 3$_{0,3}-$2$_{0,2}$ & 0.0 & 0.15 & 333.7 & I N\\ 
H$_2$CO & 3$_{2,2}-$2$_{2,1}$ & 0.0 & 0.07 & 196.6 & I P\\ 
H$_2$CO & 3$_{2,1}-$2$_{2,0}$ & 0.0 & 0.07 & 195.7 & I P\\ 
H$_2^{13}$CO & 3$_{1,2}-$2$_{1,1}$ & 0.3 & 0.52 & 16.4 & II NE\\ 
HNCO & 10$_{0,10}-$9$_{0,9}$ & 0.07 & 0.15 & 276.8 & I N\\ 
HNCO & 10$_{1,10}-$9$_{1,9}$ & 0.07 & 0.07 & 223.5 & I P\\ 
HNCO & 10$_{1,9}-$9$_{1,8}$ & 0.07 & 0.07 & 239.8 & I P\\ 
HNCO & 10$_{3,8}-$9$_{3,7}$ & 0.07 & 0.15 & 41.9 & I N\\ 
HC$_3$N & 24$-$23 & -0.07 & 0.0 & 439.7 & I P\\ 
HC$^{13}$CCN & 24$-$23 & 0.22 & -0.37 & 11.1 & II SE\\ 
HC$_3$N;$v_6$=1 & 24$-$23,l=1e & -0.22 & -0.15 & 12.0 & I SW\\ 
HC$_3$N;$v_7$=1 & 24$-$23,l=1f & -0.07 & 0.0 & 132.7 & I P\\ 
HC$_3$N;$v_7$=2 & 24$-$23,l=0 & 0.0 & -0.07 & 19.8 & I P\\ 
HC$_3$N;$v_7$=2 & 24$-$23,l=2e & -0.07 & 0.0 & 23.5 & I P\\ 
CH$_2$CO & 11$_{1,11}-$10$_{1,10}$ & 0.07 & 0.52 & 12.3 & I N\\ 
t-HCOOH & 10$_{0,10}-$9$_{0,9}$ & 0.15 & 0.15 & 48.2 & I NE\\ 
CH$_3$CN & 12$_{1}-$11$_{1}$ & 0.0 & 0.07 & 608.1 & I P\\ 
CH$_3$CN & 12$_{2}-$11$_{2}$ & 0.0 & 0.07 & 355.6 & I P\\ 
CH$_3$CN & 12$_{3}-$11$_{3}$ & 0.0 & 0.07 & 449.5 & I P\\ 
CH$_3$CN & 12$_{4}-$11$_{4}$ & 0.0 & 0.07 & 249.0 & I P\\ 
CH$_3$CN & 12$_{5}-$11$_{5}$ & 0.0 & 0.07 & 190.9 & I P\\ 
CH$_3$CN & 12$_{6}-$11$_{6}$ & 0.0 & 0.07 & 229.3 & I P\\ 
CH$_3$CN & 12$_{7}-$11$_{7}$ & 0.0 & 0.07 & 65.7 & I P\\ 
CH$_3$CN & 12$_{8}-$11$_{8}$ & 0.0 & 0.07 & 35.1 & I P\\ 
CH$_3$OH & 20$_{1,19}-$20$_{0,20}$(E) & 0.07 & 0.3 & 93.0 & III N\\ 
CH$_3$OH & 23$_{5,18}-$22$_{6,17}$(E) & 0.07 & 0.44 & 13.1 & II N\\ 
CH$_3$OH;$v_t$=1 & 6$_{1,5}-$7$_{2,5}$(A) & 0.15 & 0.3 & 138.8 & III NE\\ 
CH$_3$OH;$v_t$=1 & 15$_{6,9}-$16$_{5,11}$(A) & 0.07 & 0.3 & 13.9 & I N\\ 
NH$_2$CHO & 10$_{1,9}-$9$_{1,8}$ & 0.0 & 0.0 & 30.9 & I P\\ 
C$_2$H$_3$CN & 23$_{3,20}-$22$_{3,19}$ & -0.15 & 0.0 & 14.2 & I W\\ 
C$_2$H$_3$CN & 23$_{4,20}-$22$_{4,19}$ & 0.0 & 0.07 & 20.9 & I P\\ 
C$_2$H$_3$CN & 23$_{6,18}-$22$_{6,17}$ & 0.0 & 0.0 & 32.1 & I P\\ 
C$_2$H$_3$CN & 23$_{10,14}-$22$_{10,13}$ & 0.07 & 0.07 & 6.9 & I P\\ 
CH$_3$OCHO & 17$_{3,14}-$16$_{3,13}$(E) & 0.07 & 0.3 & 31.8 & III N\\ 
CH$_3$OCHO & 17$_{3,14}-$16$_{3,13}$(A) & 0.07 & 0.37 & 42.5 & III N\\ 
CH$_3$OCHO & 17$_{4,13}-$16$_{4,12}$(A) & 0.07 & 0.37 & 38.8 & III N\\ 
C$_2$H$_5$CN & 24$_{2,22}-$23$_{2,21}$ & -0.07 & 0.07 & 25.4 & I P\\ 
C$_2$H$_5$CN & 24$_{3,21}-$23$_{3,20}$ & -0.07 & 0.22 & 25.9 & I N\\ 
CH$_3$OCH$_3$ & 23$_{3,21}-$23$_{2,22}$(EE) & -0.22 & -0.67 & 6.7 & I SW\\ 
CH$_3$COCH$_3$ & 20$_{2,18}-$19$_{3,17}$(EE) & 0.07 & 0.3 & 14.0 & I N\\ 
CH$_3$COCH$_3$ & 21$_{1,20}-$20$_{2,19}$(AE) & 0.15 & 0.22 & 10.1 & I NE\\ 
CH$_3$COCH$_3$ & 21$_{1,20}-$20$_{1,19}$(EE) & 0.37 & 0.0 & 11.3 & II E\\ 
CH$_3$COCH$_3$ & 22$_{1,22}-$21$_{1,21}$(EE) & 0.3 & 0.15 & 17.6 & II NE\\ 
CH$_3$COCH$_3$ & 22$_{1,22}-$21$_{1,21}$(AA) & 0.22 & 0.22 & 17.9 & III NE\\ 
UL & $-$ & -0.07 & -0.07 & 32.8 & I P\\ 
\hline
\end{tabular}
\end{scriptsize}
\tablefoot{$^{(a)}$ At a distance of $3.33\,$kpc, 0\as07 is equivalent to a projected spatial scale of $230$\,au. $^{(b)}$ Classification of the spatial distribution (I: single-peaked, II: double-peaked, III: ring-like) and position of the peak integrated intensity: P for the position of the 1.37\,mm continuum peak at $\alpha$(J2000) = 20:29:24.88 and $\delta$(J2000) = +40:11:19.47; cardinal directions (e.g., N, NE) for orientation with respect to the position of the 1.37\,mm continuum peak. }
\end{table}

We investigate the spatial distribution of the molecular emission with line integrated intensity maps at high angular resolution. Figures \ref{Fig:Moment0_1} and \ref{Fig:Moment0_2} shows the integrated intensities obtained from the NOEMA spectral line data. The flux is integrated along five channels covering a width of 15\,km\,s$^{-1}$ (the average line FWHM is 5\,km\,s$^{-1}$, Sect \ref{sec:XCLASSFittPeak}) centered at the rest frequency of each transition. We carefully chose transitions where line-blending is not an issue. These maps clearly indicate chemical differentiation: a classification of the spatial distribution is summarized in Table \ref{tab:Moment0_Pos}. As we observe a similar chemical segregation as reported by \citet{Jimenez2012}, we adopt a similar classification for the morphology of the spatial distribution of the molecular emission. Type I molecules have a single-peaked distribution (spatially), while type II molecules have a double-peaked distribution (spatially). Type III molecules have a ring-like morphology. Towards the 1.37\,mm continuum emission peak, sulfur- and nitrogen-bearing species are most abundant where the density and temperature are highest (e.g., SO, OCS, SO$_2$, HC$_3$N, CH$_3$CN). Emission from most COMs is strongest in a ring-like structure towards the northeast, most visible in CH$_3$OH. The emission of CH$_2$CO is shifted $\sim$0\as4 north towards the position of VLA\,3-N which might be in a colder and younger stage. At this position, the emission of COMs is also bright but more spread out. The asymmetric emission from less abundant and optically thin isotopologues (O$^{13}$CS, H$_2^{13}$CO, HC$^{13}$CCN) may trace shocked regions where their abundances are enhanced. \citet{Palau2017} show that such asymmetries in molecular emission were also found at similar spatial scales of $\sim$600\,au towards the prototypical high-mass protostar IRAS\,20126+4104, and also that the observed asymmetries for H$_2^{13}$CO, CH$_3$OH, CH$_2$CO, and CH$_3$OCH$_3$ are consistent with a shock model coupled to a large gas-grain chemical network for a timescale of $\sim$2\,000\,years. The integrated intensity of the unidentified line (UL) at 219.616\,GHz (bottom right panel in Fig. \ref{Fig:Moment0_2}) is centered around the position of the 1.37\,mm continuum peak. 

The complex emission of various types of species shows that the formation of molecules depends heavily on the local physical conditions, such as the envelope, shocked regions, and outflows. It is very likely that the source consists of multiple chemically distinct sub-components.

\section{Chemical model}\label{sec:chemicalmodel}
The observational analysis of the AFGL\,2591 CORE data reveal a large degree of chemical complexity both in abundances and spatial distribution. In this section, we use the physical-chemical modeling code {\tt MUSCLE} \citep[MUlti Stage ChemicaL codE, a detailed description of the model is given in][]{Gerner2014} to study the formation and destruction of the observed molecular abundances. {\tt MUSCLE} computes a static radial physical structure coupled with the time-dependent chemical kinetics code {\tt ALCHEMIC} \citep{Semenov2010,Alchemic2017} including both gas-phase and grain-surface chemistry with reaction rates obtained from KIDA \citep[][]{Wakelam2012, Wakelam2015}. The computed abundance profiles are beam-convolved and thus converted into model column densities as a function of time. The model column densities are compared to the observed column densities, and with a $\chi^2$ analysis the best-fit physical structure and chemical age are determined.

The temperature $T(r)$ and the density profile $n(r)$ of the spherically symmetric model core with outer radius $r_{\mathrm{out}}$ have a constant inner part for $r < r_{\mathrm{in}}$ and obey a power-law profile for $r\geq r_{\mathrm{in}}$:
\begin{equation}
\begin{array}{l@{\qquad}c}
n(r) = n_{0} & r < r _{\mathrm{in}},\\
n(r) = n_{0} \times \bigg(\frac{r}{r_{\mathrm{in}}}\bigg)^{-p} & r_{\mathrm{in}} \leq r \leq r_{\mathrm{out}}, \\
T(r) = T_{0} & r < r _{\mathrm{in}},\\
T(r) = T_{0} \times \bigg(\frac{r}{r_{\mathrm{in}}}\bigg)^{-q} & r_{\mathrm{in}} \leq r \leq r_{\mathrm{out}}. \\
\end{array}
\end{equation}
The parameters of the physical structure ($r_{\mathrm{in}}$, $r_{\mathrm{out}}$, $T_{0}$, $p$) can either be fixed or be varied as fit parameters within given ranges.

\subsection{Initial abundances and conditions}\label{sec:ModelAssump}
The adopted {\tt MUSCLE} input parameters are summarized in Table \ref{tab:model_input}. We assume that internal sources do not generate X-rays: $\zeta_{\mathrm{X}} = 0$\,s$^{-1}$. High energetic UV and X-ray radiation emitted by the forming YSO(s) in the center can affect the envelope material of the AFGL\,2591 hot core through the outflow cavity \citep[as studied by, e.g.,][]{Benz2007}, but in our simplistic spherically symmetric model, the high energetic X-rays would be immediately absorbed by the dense gas in the center. A high extinction of 100$^{\mathrm{mag}}$ in the outer envelope shields the core from the interstellar radiation field. At infrared wavelengths, the AFGL\,2591 region seems to be isolated within 30$''$, so external heating is not important \citep{vanderTak2000}. Only cosmic rays (CRs) can penetrate into the central part of the core. We employ a cosmic ray ionization rate of 5$\times$10$^{-17}$\,s$^{-1}$ \citep{vanderTakDis2000}. We use a low value of 10$^{-5}$ for the UV photodesorption yield, as measured for CH$_3$OH \citep{CruzDiaz2016, Bertin2016}, which gives the probability that a CR-induced UV photon kicks a molecule off an icy grain.

\begin{table}[htb]
\caption{Model input parameters.}
\label{tab:model_input}
\centering
{\renewcommand{\arraystretch}{1.2}
\begin{tabular}{ll}
\hline \hline
Parameter & Value or Range\\
\hline
\textbf{Radiation field:}&\\
CR ionization rate $\zeta_{\mathrm{CR}}$&5($-17$)\,s$^{-1}$\\
X-ray ionization rate $\zeta_{\mathrm{X}}$&0\,s$^{-1}$\\
Extinction at the cloud edge $A_v$&100$^{\mathrm{mag}}$\\
UV photodesorption yield &1($-$05)\\
\hline
\textbf{Grain properties:} &\\
Grain radius $r_{\mathrm{g}}$ &0.1\,$\mu$m\\
Dust density $\rho_{\mathrm{d}}$ & 3\,g\,cm$^{-3}$\\
Gas-to-dust mass ratio $\eta$ & 150\\
Surface diffusivity $E_{\mathrm{Diff}}$/$E_{\mathrm{Bind}}$ & 0.4\\
Mantle composition & Mg$_x$Fe$_{x-1}$SiO$_4$ (olivine)\\
\hline
\textbf{Pre-HMC stages} & \\
\textbf{IRDC stage $^a$} & \\
Inner radius $r_{\mathrm{in}}$ &$12\,700$\,au\\
Outer radius $r_{\mathrm{out}}$ &$0.5$\,pc\\
Temperature $T_0$ at $r_{\mathrm{in}}$ &$11.3$\,K\\
Temperature power-law index $q$ & 0.0 (isothermal)\\
Density $n_0$ at $r_{\mathrm{in}}$ & 1.4(5)\,cm$^{-3}$\\
Density power-law index $p$ & 1.5\\
$\tau_\mathrm{IRDC}$ & $16\,500$\,years\\
\hline
\textbf{HMC stage:} & \\
Inner radius $r_{\mathrm{in}}$ &$50-700$\,au\\
Inner radius step $\Delta r_{\mathrm{in}}$ & 50\,au\\
Outer radius $r_{\mathrm{out}}$ &$7\,000-31\,000$\,au\\
Outer radius step $\Delta r_{\mathrm{out}}$ & 2\,000\,au\\
Temperature power-law index $q$ &0.41\\
Temperature $T_0$ at $r_{\mathrm{in}}$ $^b$ & $T_0 = 255\,\mathrm{K} \times \bigg(\frac{r_{\mathrm{in}}}{691\,\mathrm{au}}\bigg)^{-q}$ \\
Density power-law index $p$ &$1.0-2.5$\\
Density power-law index step $\Delta p$ &0.1\\
$\tau_{\mathrm{HMC}}$ & $10-10^5$\,years \\
$\Delta \tau_{\mathrm{HMC}}$ & 100 logarithmic steps\\
\hline
\end{tabular}
}
\tablefoot{a(b) = a$\times$10$^{\mathrm{b}}$. $^{ (a)}$ Taken from \citet{Gerner2015}. $^{ (b)}$ Constraint from the observations described in Sect. \ref{subsec:T_profile}.}
\end{table}

Grain-surface reactions depend on the geometrical structure of dust particles \citep{Gerin2013}. We consider spherical grains with an amorphous (unstructured) silicate composition. Each dust particle has 1.88$\times$10$^6$ surface sites \citep{Semenov2010} where species can freeze out and chemical reactions can occur via the Langmuir-Hinshelwood mechanism \citep{Hasegawa1992, Biham2001}. Neutral species and electrons can stick to the grains with a probability of 100\,\%. Only hydrogen atoms are allowed to tunnel through the surface barriers. After a reaction occurs, products can directly desorb from the grain with a probability of 1\,\% \citep{Garrod2007,Vasyunin2013}. In addition, species can desorb by thermal processes, CRs, and UV radiation. Self-shielding of H$_2$ and CO is included as well. For consistency with our observational analysis, we use a gas-to-dust mass ratio of 150 \citep{Draine2011}. The diffusivity of species on dust grains can be parameterized by the ratio of the diffusion energy to the binding energy and as it is poorly constrained, we employ an experimentally motivated value of 0.4 \citep{Cuppen2017}.

The evolutionary past of the physical structure and chemical composition of AFGL\,2591 is not known. As initial abundances, we use the best-fit IRDC abundances from the study by \citet{Gerner2015}. For 19 well-studied IRDCs, column densities of 18 molecules including deuterated species were obtained using IRAM\,30\,m observations at 1 and 3\,mm. The column densities of the molecules were averaged and modeled to create abundances of a template IRDC. These authors derive an IRDC chemical age of $\tau_{\mathrm{IRDC}} = 16\,500$\,years using {\tt MUSCLE} \citep[Table A.4 and A.8 in][]{Gerner2015}.

Finally, the physical structure of the AFGL\,2591 hot core has to be defined. Our NOEMA data are missing the short-spacings in the continuum yielding a missing flux of 84\,\%, so we can only determine a lower limit for the outer radius of 7\,000\,au (Fig. \ref{Fig:Cont}). From the observed radial temperature profile shown in Fig. \ref{Fig:TMap}(b), we can adopt an upper limit for the inner radius of 700\,au which is half a beam size. The density index $p$ is varied from 1.0 to 2.5. The temperature profile is fixed to the radial profile derived from the observations presented in Sect. \ref{subsec:T_profile} (Eq. \ref{eq:obsT}) and thus for each physical model the temperature $T_0$ at the inner radius is calculated according to $T_0 = 255\,\mathrm{K} \times (\frac{r_\mathrm{in}}{691\,\mathrm{au}})^{-0.41}$. In total, we run the chemical code {\tt ALCHEMIC} on 2912 physical models ($\#r_{\mathrm{in}}\times \#r_{\mathrm{out}}\times \#p = 14 \times 13 \times 16$). For each model, it takes $\sim$75\,s using 32 cores to compute the chemical evolution with {\tt ALCHEMIC} up to 100\,000\,years on 40 radial grid points.
\subsection{Best-fit model}\label{bestfitmodel}
Using a $\chi^2$ analysis, the observed column densities $N_\mathrm{obs}$ are compared to the computed model column densities $N_\mathrm{mod}$ for each physical model and time step. \citet{Vasyunin2004} studied the error propagation of the abundances of chemical models due to the large uncertainties and approximations of rate coefficients. These authors estimated an error of $0.5-1.5$ orders of magnitude depending on the size of the molecule. \citet{Wakelam2005} found that for older hot cores with $\tau > 10^4$\,years, the uncertainties in the chemical model can be large for some species and therefore care should be taken when comparing observed and modeled abundances. We therefore assume that a molecule is modeled well if the modeled and observed column densities agree within one order of magnitude\footnote{In the current {\tt MUSCLE} version, uncertainties of the observed column densities cannot yet be included as an additional constraint.}.

\begin{table}[htb]
\caption{Best-fit model of the AFGL\,2591 VLA\,3 hot core.}
\label{tab:bestfitmodel}
\centering
\begin{footnotesize}
\begin{tabular}{lc}
\hline \hline
 & Best-fit Model Parameters \\
\hline
Inner radius $r_{\mathrm{in}}$ & 700\,au \\
Outer radius $r_{\mathrm{out}}$ &$29\,000$\,au\\
Temperature power-law index $q$ & $0.41$ \\
Temperature $T_0$ at $r_{\mathrm{in}}$ & $253.7$\,K \\
Density power-law index $p$ & $1.0$ \\
Density $n_0$ at $r_{\mathrm{in}}$ & $1.7(7)$\,cm$^{-3}$ \\
Mass $M$ & 487\,$M_\odot$ \\
HMC stage $\tau_{\mathrm{HMC}}$ & 4\,642\,years\\
Well fitted molecules & 10/14 \\
 $\chi^2$ & 0.5295\\
\hline
\end{tabular}
\end{footnotesize}
\tablefoot{a(b) = a$\times$10$^{\mathrm{b}}$.}
\end{table}

\begin{figure}[htb]
\resizebox{\hsize}{!}{\includegraphics{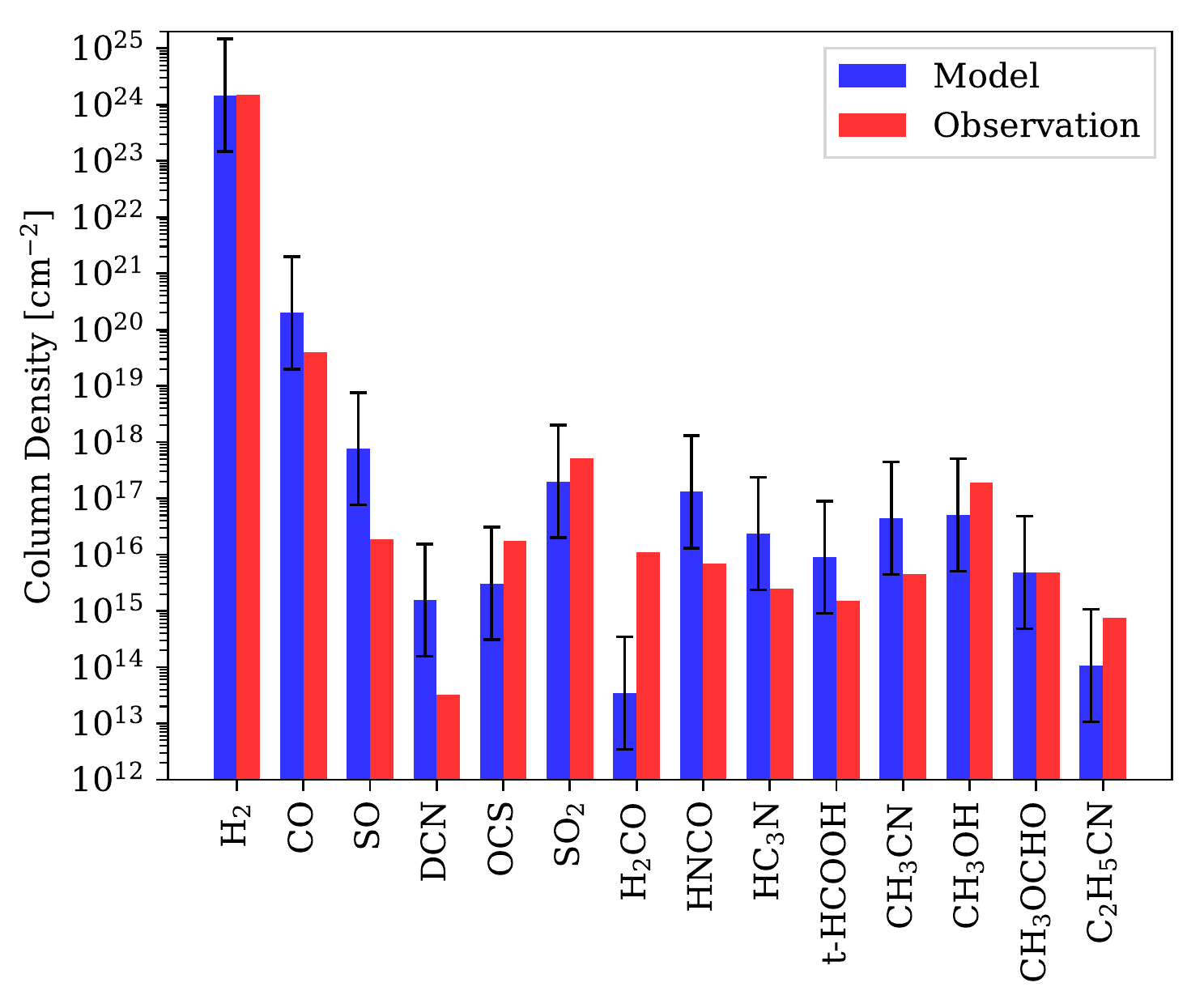}}
\caption{Comparison of the observed (red) and modeled (blue) column densities.}
\label{Fig:Model_Results_Bar}
\end{figure}

\begin{figure*}
\centering
\includegraphics[width=17cm]{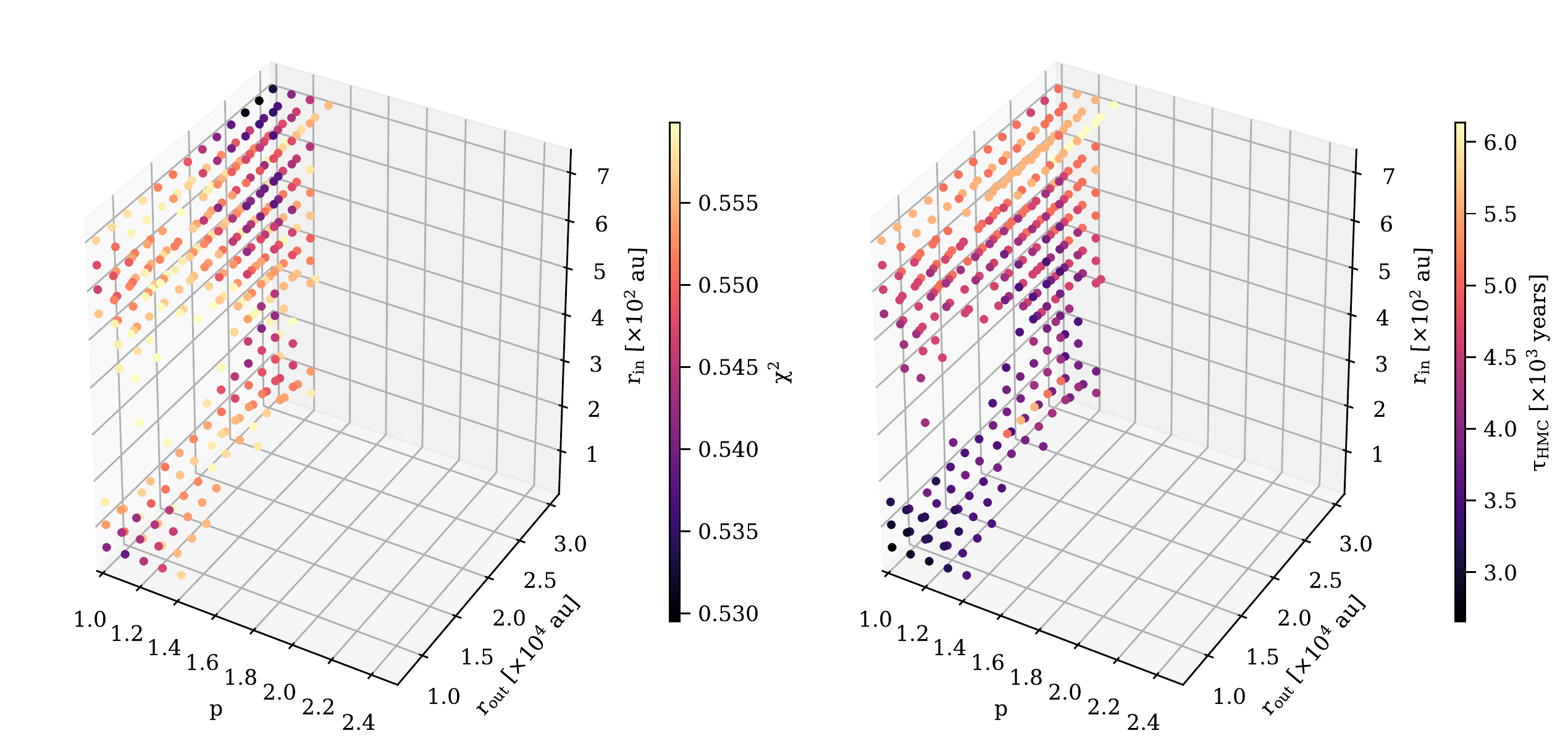}
\caption{{\tt MUSCLE} fit parameter space. Left panel: Physical models with $\chi^2 < 0.56$. Right panel: Best-fit chemical age $\tau_\mathrm{HMC}$ of the physical models with $\chi^2 < 0.56$.}
\label{Fig:ModelParamSpace}
\end{figure*}

\begin{figure*}
\centering
\includegraphics[width=17cm]{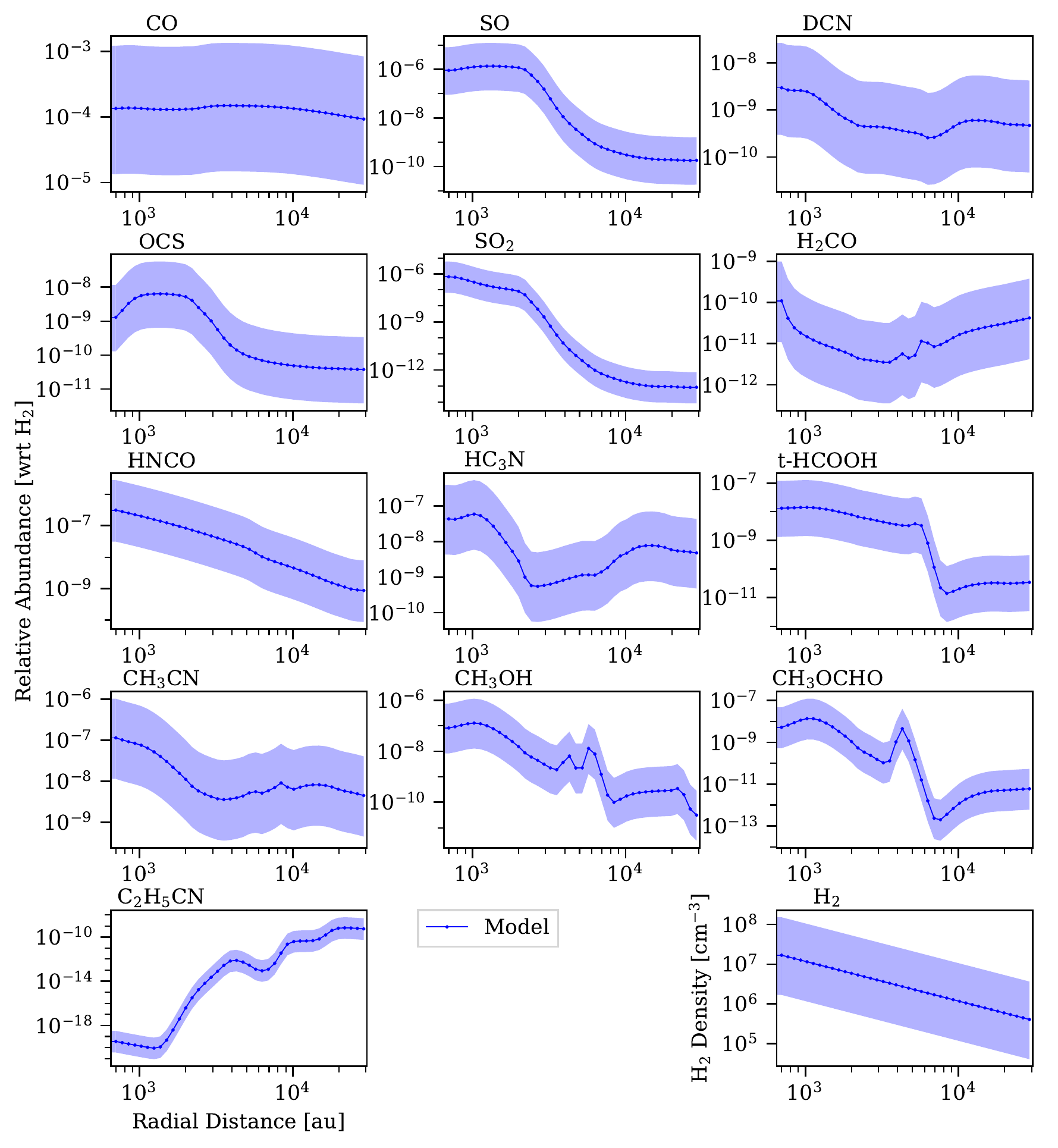}\\
\caption{Radial abundance profiles of the {\tt MUSCLE} best-fit model, relative to the radial H$_2$ density profile in the bottom right panel.}
\label{Fig:ModelRadialProfile}
\end{figure*}

The column densities derived from the observations given as input are listed in Table \ref{tab:ColDensT}. The H$_2$ column density is calculated from the 1.37\,mm continuum emission at the peak position, as analyzed in Sect \ref{subsec:h2gas}. The CO column density is inferred from the observed C$^{18}$O abundance assuming an isotopic ratio of $^{16}$O/$^{18}$O = 500 \citep{Wilson1994}. For molecules whose lines from the vibrational ground state and from vibrationally excited states were observed, an average value for the column density is given as input.
 
The best-fit parameters from the model with the lowest $\chi^2$ are summarized in Table \ref{tab:bestfitmodel}. The best-fit model has a hot core lifetime $\tau_{\mathrm{HMC}} \approx 4\,600$\,years, with a density power-law index of $p = 1.0$. A comparison between the modeled and observed column densities is shown in Fig. \ref{Fig:Model_Results_Bar}. For 10 out of 14 molecules the modeled and observed column densities agree within the uncertainties. For SO, DCN, and HNCO the model column density is too high compared to the observations with $\frac{N_\mathrm{model}}{N_\mathrm{obs}} = 40$, 48, and 18, respectively, while for H$_2$CO the modeled column density is too low with $\frac{N_\mathrm{model}}{N_\mathrm{obs}} = 3.1 \times 10^{-3}$. \citet{Palau2017} found that there is an enhancement of gas-phase H$_2$CO sputtered off the grains along an outflow of the high-mass protostar IRAS 20126+4104. Thus, one explanation for the underestimation of the H$_2$CO abundance in our model could be that our chemical model does not include shock chemistry. The integrated intensity of H$_2^{13}$CO (Fig. \ref{Fig:Moment0_1}) shows an enhancement along a potentially blueshifted NE outflow part as traced by SO (Fig. \ref{Fig:Outflow}).

The $\chi^2$ and $\tau_\mathrm{HMC}$ distribution in the ($r_{\mathrm{in}}\times r_{\mathrm{out}}\times p$)-parameter space is shown in Fig. \ref{Fig:ModelParamSpace} for all models with $\chi^2 < 0.56$. The best-fit model lies in a parameter space region with $p \approx 1.0 - 1.4 $ and $\tau_\mathrm{HMC} \approx 3\,000 - 6\,000$\,years. With {\tt MUSCLE} we cannot constrain $r_{\mathrm{in}}$ or $r_{\mathrm{out}}$ as within our model parameter space all values reach a small $\chi^2$ (Fig. \ref{Fig:ModelParamSpace}).

Within the best-fit model it is possible to analyze the radial variation of the modeled abundances, which is shown in Fig. \ref{Fig:ModelRadialProfile} at the time of the best-fit chemical age $\tau_\mathrm{HMC}$. The modeled CO, DCN, and H$_2$CO abundances stay approximately constant. The sulfur-bearing species SO, OCS, and SO$_2$ have a high abundance in the center up to $\sim$4\,000\,au ($> 124$\,K), with a sharp drop afterwards. The nitrogen-bearing species HNCO, HC$_3$N, and CH$_3$CN are enhanced in the central region up to 3\,000\,au ($> 140$\,K) as well. A similar profile is seen for t-HCOOH, CH$_3$OH, and CH$_3$OCHO, but the abundances drop at 6\,000\,au ($> 100$\,K). In contrast, the abundance of C$_2$H$_5$CN rises only towards the outer region $> 10\,000$\, au ($> 85$\,K). For C$_2$H$_5$CN, higher model temperatures would be required in order to form it efficiently in the densest inner region.

With a hot core stage timescale of $\tau_{\mathrm{HMC}} \approx 4\,600$\,years, the total lifetime ($\tau = \tau_{\mathrm{IRDC}} + \tau_{\mathrm{HMC}}$) of AFGL\,2591 VLA\,3 is 21\,100\,years. This is in rough agreement with previous studies that determined the chemical age of the hot core $\tau$ of a few $10^4$\,years: $\sim$30\,000\,years \citep{Doty2002}, $\sim$50\,000\,years \citep{Staeuber2005}, $\sim$80\,000\,years \citep{Doty2006}, and $10\,000-50\,000$\,years \citep{KazmierczakBarthel2015}. Theoretical models of the formation of massive stars also predict lifetimes $< 10^5$\,years for luminous YSOs \citep{Davies2011,Mottram2011}.
\section{Discussion}
\subsection{Observed and modeled density and temperature structure of the AFGL\,2591 hot core}\label{sec:disphysstruc}

\begin{table*}[htb]
\begin{scriptsize}
\centering
\caption{Literature comparison of the density power-law index $p$.}
\label{tab:pqcomparison}
\begin{tabular}{llcll}
\hline \hline
Reference & Source(s) & $p$ & Method & Spatial Scale $^{a}$\\
\hline
\citet{vanderTak2000} &12 massive young stars & $1.0-1.5$& CS line and continuum modeling & clump\\
&2 hot core types & 2.0 & CS line and continuum modeling & clump\\
\citet{Choi2000} & Mon R2 dense core & 0.9 & CS line Large Velocity Gradient simulations & clump \\
& Mon R2 dense core & 0.8 & CS line radiative transfer simulations & clump\\
\citet{Beuther2002} & 69 massive star-forming regions &$1.6\pm0.5$ & continuum radial intensity profile modeling & clump\\
\citet{Mueller2002} & 31 dense cores &$1.8\pm0.4$ & continuum SED and radial intensity profile modeling & clump\\
\citet{Hatchell2003} &10 UC H{\sc ii} regions &$1.25-2.25$& continuum SED and radial intensity profile modeling + CS line modeling & clump\\
\citet{Chen2006} & W3(H2O) hot core & 1.52 & continuum radiative transfer model & core\\
\citet{Beuther2007B} &3 cores in IRAS05358+3543 &$1.5-2$& continuum $uv$-plane analysis & core\\
\citet{Zhang2009} &2 high-mass cores in G28.34+0.06&$1.6-2.1$& continuum $uv$-plane analysis & core \\
\citet{Longmore2011} & massive protocluster G8.68$-$0.37 &$1.8\pm0.2$& continuum $uv$-plane analysis & core \\
\citet{Palau2014} &19 massive dense cores &$1.5-2.4$& continuum SED and radial intensity profile model & clump + core\\
\citet{Beltran2018} & G31.41+0.31 hot core &$2.2$& radial H$_2$ column density profile fit & core\\
This work & AFGL\,2591 hot core & $1.7\pm0.1$ & continuum $uv$-plane analysis & core\\
\hline
\end{tabular}
\end{scriptsize}
\tablefoot{$^{(a)}$ Based on the angular resolution of the observations ($> 10''$ for ``clump'' and $< 10''$ for ``core'' scales).}
\end{table*}

In Sect. \ref{sec:physenv} we present the analysis of the physical environment of the hot core based on the CORE observational data. First, we inferred the kinetic temperature structure using H$_2$CO and CH$_3$CN and derived the temperature power-law index $q = 0.41\pm0.08$ assuming spherical symmetry (Fig. \ref{Fig:TMap}), where the temperature profile may not follow a single power-law. Furthermore, using the visibilities of the 1.37\,mm continuum emission (Fig. \ref{Fig:VisibilityFit}), we inferred the density power-law index $p = 1.7 \pm 0.1$ of the source. Again, spherical symmetry was assumed for the analysis. In reality, this assumption is most likely not valid for both the temperature and density profile due to the outflow and the potential fragmentation on even smaller scales, as discussed in Sects. \ref{sec:physenv} and \ref{sec:moment0}. However, the assumption of spherical symmetry is sufficient to describe the surrounding envelope.

Deriving the density structure in high-mass star-forming regions has not been attempted frequently, but a few studies have been put forward in previous years. Table \ref{tab:pqcomparison} summarizes the results for the density profiles $p$ obtained in other studies. There are several theoretical models that explain the collapse of a gas cloud into a core. Numerical calculations of a collapsing spherically symmetric cloud by \citet{Larson1969} have shown that without initial pressure gradients the free-fall collapse is non-homologous and the density profile approaches $p = 2.0$. \citet{Shu1977} studied the collapse of an isothermal cloud that is in hydrostatic equilibrium. The initial density distribution follows a power-law with $p = 2.0$ and the accretion rate is constant during the collapse. \citet{McLaughlin1997} refined this simple model, considering a non-isothermal gas distribution and a logotropic equation of state, which can best explain observations of both low- and high-mass cores from small to cloud scales. In this case, an initial value of $p = 1.0$ holds and during the collapse the accretion rate increases. In both models the gas eventually free-falls onto the center with $p = 1.5$. As summarized in Table \ref{tab:pqcomparison}, observational determinations of the density profile $p$ range from $1 - 2$. One explanation for such large variations could be that during the evolution in HMSF the temperature and density structure may change drastically, especially when considering different formation and accretion mechanisms that require non-spherical symmetry. Furthermore, depending on the tracer and the spatial resolution, different spatial scales and regions are traced, which might have different density profiles and asymmetries.

Regarding the density profile of the AFGL\,2591 hot core, our model predicts $p < 1.4$, in contrast to our observational value of $p = 1.7\pm0.1$. We cannot determine whether the density profile is dominated by non-thermal processes ($p = 1$), free-fall collapse ($p = 1.5$), or thermal processes ($p = 2$). The observed temperature profile presented in Fig. \ref{Fig:TMap} is not fully spherically symmetric and does not obey a single power-law. As a consequence, the model temperature in the outer region is higher than observed. As $q$ and $p$ are correlated, a steeper temperature index results in a flatter density index. Mottram et al. (subm.) and \citet{Beltran2018} observe steeper values for hot cores: $q = 0.7$ and $q = 0.8$, respectively. The density profile of AFGL\,2591 was derived in a previous study based on single-dish observations: \citet{vanDerTak1999} modeled CS and C$^{34}$S emission lines to infer a density power-law index of $1.25\pm0.25$. Our observations of AFGL\,2591 VLA\,3 at high angular resolution suggest that $p > 1.6$. These differences indicate that the inner core may have a steep density profile, while the density profile of the large-scale envelope may be flatter.
\subsection{Model uncertainties}\label{sec:modeluncert}
The observed molecular column densities presented in Sect. \ref{sec:XCLASSFittPeak} were applied to the physical-chemical 1D model {\tt MUSCLE} in order to study the physical structure of the hot core and radial profiles of the molecular abundances. The observed molecular emission around the hot core (shown in Fig. \ref{Fig:Moment0_1} and Fig. \ref{Fig:Moment0_2}) reveal that on subarcsecond scales the structure of the molecular gas is highly asymmetric. Therefore, the modeling approach of a spherically symmetric core is too simplistic to understand all the chemical properties on such small scales. For instance, dense gas may be shielded by the inner disk, molecules in the outflow cavity might be strongly affected by high-energy radiation of the forming protostar. Unfortunately, a full 3D physical-chemical model is not yet available to investigate these effects further.

While {\tt MUSCLE} is able to model several evolutionary stages of HMSF, within each stage the physical structure is static. There is no temperature and density evolution within a model stage, but in reality the core is collapsing, where the accretion disk and outflow evolve as well. The reaction rates are very sensitive to temperature and density changes and the high-temperature structure that we observe was preceded by a warm-up phase. We have no knowledge of the conditions of the pre-HMC stages of AFGL\,2591. Therefore, the best-fit IRDC abundances from \citet{Gerner2015} were applied as a starting point for the HMC stage of AFGL\,2591. We tested different initial conditions of the abundances based on previous {\tt MUSCLE} studies: using the IRDC+HMPO stage from \citet{Gerner2015}, the model performs worse with fewer molecules modeled well and a higher $\chi^2$ value. In that case the chemical age would be $> 49\,000$\,years suggesting that the hot core is younger. Using the IRDC + HMPO stage from the CORE pilot region NGC\,7538\,S \citep{Feng2016}, we obtain similar results, but a steeper density profile. However, the modeled physical structure is much more compact with an outer radius of 1\,100\,au and an extrapolation of the chemical abundances to our AFGL\,2591 physical models with much higher outer radii is not reasonable. Many molecules, including simple COMs and their precursors readily form during the cold IRDC phase \citep[e.g.,][]{Vasyunina2014}. Hence the initial atomic and molecular abundances play an important role. As studied by \citet{Wilson1994} and \citet{Wannier1980}, among others, the abundances in the ISM depend on the local environment and there are gradients from the Galactic center towards the outer regions, as high-mass stars quickly enrich the local environment with heavy metals. Thus, it is necessary to study the chemical conditions in the ISM, the influence of the environment, and the evolutionary timescales in HMSFRs with a large statistical sample at a high spatial resolution, analogous to the work by \citet{Gerner2014, Gerner2015}, among others, who used single-dish observations. A study of the chemical content of all 18 regions at high angular resolution within the large program CORE will be presented in a future publication (Gieser et al. in prep.).

One uncertainty of the model is the assumption of the radiation field. During HMSF, when the star reaches the main sequence, it is still deeply embedded in the cloud and envelope. Therefore, the central source may already emit strong UV radiation and X-rays, which have a strong impact on the chemistry. This enhances the formation of ions but also the destruction of large molecules. The high-energy radiation can escape the central part of the source through the outflow and winds. Numerical simulations by \citet{Kuiper2018} have shown that photoionization feedback plays an important role in the outflows of massive stars on scales from 0.01$-$1 pc. Previous studies of AFGL\,2591 have shown that some observed molecular abundances can only be explained when a central heating source is included in the model \citep{Benz2007, Bruderer2009}. The chemical segregation observed by \citet{Jimenez2012} was also explained by molecular UV photo-dissociation and high-temperature gas-phase chemistry in the central region. In addition, as recent observations by \citet{Csengeri2018} suggest, the physical structure of high-mass envelopes can be impacted by additional processes, such as shocks due to accretion.

Grain-surface reactions play an important role in the formation of molecules in the ISM \citep[e.g.,][]{Potapov2017}. In our model, we assumed spherical silicate dust grains with uniform size, but we neglect carbonaceous grains which may influence chemical reactions on the grains. Observations have revealed that dust grains follow a size distribution \citep{Mathis1977, Weingartner2001}. \citet{Simpson2013} have shown that the dust grains around AFGL\,2591 consist of both spherical and elongated grains. Both the grain size and form distribution define the effective surface area and, for larger areas, chemical reactions are more likely to occur.

In deriving the observed column densities and in the {\tt MUSCLE} model, we assumed that AFGL\,2591\,VLA\,3 consists of one hot core. Previous studies have claimed the presence of a second source (VLA\,3-N) $\sim$0\as4 towards the north \citep{Trinidad2003,Trinidad2013,Sanna2012}. We observe several outflow directions in different outflow tracers as discussed in Sect. \ref{subsec:outflow} and the chemical segregation of the molecular distribution shown in Fig. \ref{Fig:Moment0_1} and Fig. \ref{Fig:Moment0_2} indicates fragmentation on scales smaller than $1\,400$\,au. Therefore, the observed molecular emission may come from several YSOs which we do not resolve. However, with our simplistic model we are still able to capture the most essential physics and chemistry of simple gas-phase species.

\section{Conclusions} \label{section:conclusions}

In this study, we analyzed continuum and spectral line data of AFGL\,2591\,VLA\,3 at 1.37\,mm obtained within the NOEMA large program CORE. We derived the radial temperature and density profiles, studied the outflow and determined the molecular abundances towards the position of the 1.37\,mm continuum peak. We modeled the chemical abundances using the physical-chemical code {\tt MUSCLE} and found a rough agreement with previous studies for the chemical age of the hot core. Our main conclusions of this study are the following:

\begin{itemize}
\item The 1.37\,mm spectra reveal a large chemical complexity in O-, N-, and S-bearing species. In contrast to previous line surveys of the hot core at lower angular resolution, the spectra of the CORE data show that AFGL\,2591 VLA\,3 indeed has a typical hot core spectrum with many molecular emission lines. We found several isotopologues containing $^{13}$C, D, $^{18}$O, $^{33}$S, and $^{34}$S and observed pure rotational ($v$=0), vibrational (up to $v$=3), and torsional ($v_t$=1) transitions.

\item The molecular thermometers CH$_3$CN and H$_2$CO were used to derive the temperature structure. We obtain a power-law index of $q = 0.41 \pm 0.08$. While the observed CH$_3$CN lines ($E_\mathrm{u}$/$k_\mathrm{B} = 183$, 247, and 326\,K) trace hot and compact emission, the distribution of the observed H$_2$CO emission lines ($E_\mathrm{u}$/$k_\mathrm{B} = 21$ and 68\,K) are more extended tracing the low-temperature gas. By combining both temperature profiles, we were able to determine the kinetic temperature over a wide range of radii from $700$ to $4\,000$\,au.

\item The visibilities of the 1.37\,mm continuum emission were used to derive the density power-law index. Assuming spherical symmetry and optically thin emission, we obtain a value of $p = 1.7 \pm 0.1$. We derived the H$_2$ column density from the 1.37\,mm continuum emission and estimated the gas mass reservoir within the inner 10\,000\,au around the continuum peak to $6.9 \pm 0.5$\,$M_{\odot}$. This is a lower limit, as the NOEMA data miss a large amount of flux due to spatial filtering.

\item The emission of CO, SiO, and SO using single-dish and interferometric observations were used to study the outflow structure. Two potential blueshifted outflow directions can be identified, one towards the west and one towards the northeast. This suggests that there could be several fragmented cores on spatial scales $< 1\,400$\,au or that we might be observing a disk wind.

\item Using {\tt XCLASS} the column density $N$ and rotation temperature $T_{\mathrm{rot}}$ of all detected molecules were determined. Some molecules lie in an outer, colder envelope such as DCN. This is also observed in another CORE target region NGC\,7538\,IRS9 (Feng et al. in prep.). Molecules such as CH$_3$CN trace the central high-temperature region. Integrated intensity maps of detected molecular lines reveal that sulfur- and nitrogen-bearing molecules have the strongest emission towards the 1.37\,mm continuum peak (e.g., SO, OCS, HC$_3$N, CH$_3$CN, NH$_2$CHO, C$_2$H$_3$CN, C$_2$H$_5$CN). Complex molecules such as CH$_3$OH, CH$_3$OCHO, and CH$_3$COCH$_3$ peak towards the north and northeast with an asymmetric ring-like structure. The integrated intensity of less abundant isotopologues (O$^{13}$CS, H$_2^{13}$CO, HC$^{13}$CCN) may trace shocked regions. This chemical segregation can be explained by the fact that the formation of specific molecules is confined to different temperature and density regions (disk, outflow, envelope) with different dynamics (e.g., shocks), and that several fragmented cores may be hidden in the central part of the source.

\item We applied the physical-chemical model {\tt MUSCLE}, including grain-surface and gas-phase chemistry, to the observed abundances of the hot core in order to derive the physical structure and chemical age. Our best-fit model can explain 10 out of 14 detected molecular species within the uncertainties with a density power-law index of $p_{\mathrm{model}} = 1.0$. The hot core chemical age is $\tau \approx 21\,100$\,years consistent with previous studies. However, keeping the complicated structure of AFGL\,2591 at scales $< 10\,000$\,au traced by the gas in mind, the assumption of spherical symmetry is too simplistic. With a static physical structure in each stage, molecules may not be modeled well due to the lack of a warm-up stage in {\tt MUSCLE}.
\end{itemize}

The purpose of this work was to use the hot core AFGL\,2591 VLA\,3 as a case study in order to investigate the chemical abundance within the CORE spectral setup at 1.37\,mm. The high spatial resolution and sensitivity obtained by NOEMA revealed a rich chemical complexity both in abundance and spatial distribution. The physical-chemical code {\tt MUSCLE} is sufficient to model a large fraction of the observed molecular abundances obtained by interferometric observations at high angular resolution assuming spherical symmetry.

\begin{acknowledgements}
The authors would like to thank the anonymous referee whose comments helped improve the clarity of this paper. This work is based on observations carried out under project number L14AB with the IRAM NOEMA Interferometer and the IRAM\,30\,m telescope. IRAM is supported by INSU/CNRS (France), MPG (Germany), and IGN (Spain). C.G., H.B., A.A., and J.C.M. acknowledge support from the European Research Council under the Horizon 2020 Framework Programme via the ERC Consolidator Grant CSF-648505. D.S. acknowledges support from the Heidelberg Institute of Theoretical Studies for the project ``Chemical kinetics models and visualization tools: Bridging biology and astronomy.'' R.K. acknowledges financial support via the Emmy Noether Research Group on Accretion Flows and Feedback in Realistic Models of Massive Star Formation funded by the German Research Foundations (DFG) under grant no. KU 2849/3-1 and KU 2849/3-2. A.P. acknowledges financial support from UNAM-PAPIIT IN113119 grant, M\'exico. A.S.M. acknowledges funding by the Deutsche Forschungsgemeinschaft (DFG) via the Sonderforschungsbereich SFB~956 Conditions and Impact of Star Formation (subproject A6). I.J.-S. acknowledges partial support by the MINECO and FEDER funding under grants ESP2015-65597-C4-1 and ESP2017-86582-C4-1-R. This research made use of Astropy\footnote{\url{http://www.astropy.org}}, a community-developed core Python package for Astronomy \citep{Astropy2013, Astropy2018}. 
\end{acknowledgements}
\bibliographystyle{aa} 
\bibliography{bibliography} 
\begin{appendix}
\section{Transition properties of detected molecular lines}\label{sec:TransProps}
Table \ref{tab:tansition_props} shows the transition properties taken from the CDMS and JPL databases of all detected spectral lines in the average AFGL\,2591 VLA\,3 spectrum presented in Fig. \ref{Fig:LineID}. The critical density is calculated based on Einstein coefficients $A_{\mathrm{ul}}$ and collisional rate coefficients $C_{\mathrm{ul}}$ from the Leiden Atomic and Molecular Database\footnote{\url{https://home.strw.leidenuniv.nl/~moldata/}}\citep[LAMDA,][]{LAMDA}: via the approximation $n_\mathrm{crit} = \frac{A_{\mathrm{ul}}}{C_{\mathrm{ul}}}$ \citep{Shirley2015}.

Vibrationally excited states are marked by v$_{\mathrm{x}}$ (with x=1, 2, 3,...). For CH$_3$OH, torsional transitions (rotation of the methyl group within the molecule) are detected and marked with v$_{t}$. If nothing is added next to the molecule name, the transition is pure rotational ($v$=0).
The quantum numbers are given depending on the symmetry of the molecule: $J^{\mathrm{upper}}-J^{\mathrm{lower}}$ for linear molecules, $J^{\mathrm{upper}}_{K}-J^{\mathrm{lower}}_{K}$ for symmetric top molecules, and $J^{\mathrm{upper}}_{K_A,K_C}-J^{\mathrm{lower}}_{K_A,K_C}$ for asymmetric top molecules.\\

\tablecaption{Transition properties (rest frequency, species, quantum numbers, upper state energy, critical density) of detected molecular lines in the average AFGL\,2591 spectrum shown in Fig. \ref{Fig:LineID}.}
\label{tab:tansition_props}

\begin{scriptsize}
\begin{center}
\tablefirsthead{
\hline \hline
Frequency & Molecule & Transition & $E_\mathrm{u}$/$k_\mathrm{B}$ & $n_{\mathrm{crit}}$ \\
(GHz) & & & (K) & (cm$^{-3}$)\\
\hline
}
\tablehead{
\hline \hline
Frequency & Molecule & Transition & $E_\mathrm{u}$/$k_\mathrm{B}$ & $n_{\mathrm{crit}}$ \\
(GHz) & & & (K) & (cm$^{-3}$)\\
\hline
}
\begin{supertabular}{lllll}
217.239 & DCN & 3$-$2 & 21 & $-$\\
217.299 & CH$_3$OH;$v_t$=1 & 6$_{1,5}-$7$_{2,5}$(A) & 374 & $-$\\
217.399 & HC$^{13}$CCN & 24$-$23 & 130 & $-$\\
 & $^{13}$CH$_3$OH & 10$_{2,8}-$9$_{3,7}$(A) & 162 & $-$\\
217.643 & CH$_3$OH;$v_t$=1 & 15$_{6,9}-$16$_{5,11}$(A) & 746 & $-$\\
 & CH$_3$OH;$v_t$=1 & 15$_{6,10}-$16$_{5,12}$(A) & 746 & $-$\\
217.832 & $^{33}$SO & 6$_{5}-$5$_{4}$ & 35 & $-$\\
217.887 & CH$_3$OH & 20$_{1,19}-$20$_{0,20}$(E) & 508 & $-$\\
218.127 & CH$_3$COCH$_3$ & 20$_{2,18}-$19$_{3,17}$(EE) & 119 & $-$\\
 & CH$_3$COCH$_3$ & 20$_{3,18}-$19$_{3,17}$(EE) & 119 & $-$\\
 & CH$_3$COCH$_3$ & 20$_{2,18}-$19$_{2,17}$(EE) & 119 & $-$\\
 & CH$_3$COCH$_3$ & 20$_{3,18}-$19$_{2,17}$(EE) & 119 & $-$\\
218.199 & O$^{13}$CS & 18$-$17 & 99 & $-$\\
218.222 & H$_2$CO & 3$_{0,3}-$2$_{0,2}$ & 21 & 2.18(06)\\
218.281 & CH$_3$OCHO & 17$_{3,14}-$16$_{3,13}$(E) & 100 & $-$\\
218.298 & CH$_3$OCHO & 17$_{3,14}-$16$_{3,13}$(A) & 100 & $-$\\
218.325 & HC$_3$N & 24$-$23 & 131 & 4.56(06)\\
218.390 & C$_2$H$_5$CN & 24$_{3,21}-$23$_{3,20}$ & 140 & $-$\\
218.402 & C$_2$H$_3$CN & 23$_{6,18}-$22$_{6,17}$ & 204 & $-$\\
 & C$_2$H$_3$CN & 23$_{6,17}-$22$_{6,16}$ & 204 & $-$\\
218.440 & CH$_3$OH & 4$_{2,3}-$3$_{1,2}$(E) & 45 & 1.20(08)\\
218.451 & C$_2$H$_3$CN & 23$_{5,19}-$22$_{5,18}$ & 180 & $-$\\
 & C$_2$H$_3$CN & 23$_{5,18}-$22$_{5,17}$ & 180 & $-$\\
218.459 & NH$_2$CHO & 10$_{1,9}-$9$_{1,8}$ & 61 & $-$\\
 & HC$_3$N;$v_5$=1/$v_7$=3 & 24$-$23,$v_5$=1$^{1+}$ & 1085 & $-$\\
218.476 & H$_2$CO & 3$_{2,2}-$2$_{2,1}$ & 68 & 2.03(06)\\
218.492 & CH$_3$OCH$_3$ & 23$_{3,21}-$23$_{2,22}$(EE) & 264 & $-$\\
218.520 & C$_2$H$_3$CN & 23$_{10,14}-$22$_{10,13}$ & 341 & $-$\\
 & C$_2$H$_3$CN & 23$_{10,13}-$22$_{10,12}$ & 341 & $-$\\
218.574 & C$_2$H$_3$CN & 23$_{4,20}-$22$_{4,19}$ & 160 & $-$\\
218.683 & HC$_3$N;$v_6$=1 & 24$-$23,l=1e & 849 & $-$\\
218.760 & H$_2$CO & 3$_{2,1}-$2$_{2,0}$ & 68 & 2.08(06)\\
218.861 & HC$_3$N;$v_7$=1 & 24$-$23,l=1e & 452 & $-$\\
 & HC$_3$N;$v_6$=1 & 24$-$23,l=1f & 849 & $-$\\
218.875 & $^{33}$SO$_2$ & 22$_{2,20}-$22$_{1,21}$ & 252 & $-$\\
218.882 & $^{33}$SO$_2$ & 22$_{2,20}-$22$_{1,21}$ & 252 & $-$\\
218.903 & OCS & 18$-$17 & 100 & 3.89(05)\\
218.981 & HNCO & 10$_{1,10}-$9$_{1,9}$ & 101 & 7.78(06)\\
218.996 & SO$_2$;$v_2$=1 & 20$_{2,18}-$19$_{3,17}$ & 971 & $-$\\
219.174 & HC$_3$N;$v_7$=1 & 24$-$23,l=1f & 452 & $-$\\
219.220 & CH$_3$COCH$_3$ & 21$_{1,20}-$20$_{2,19}$(AE) & 122 & $-$\\
 & CH$_3$COCH$_3$ & 21$_{1,20}-$20$_{1,19}$(EA) & 122 & $-$\\
 & CH$_3$COCH$_3$ & 21$_{2,20}-$20$_{2,19}$(EA) & 122 & $-$\\
 & CH$_3$COCH$_3$ & 21$_{2,20}-$20$_{1,19}$(AE) & 122 & $-$\\
219.242 & CH$_3$COCH$_3$ & 21$_{1,20}-$20$_{1,19}$(EE) & 122 & $-$\\
 & CH$_3$COCH$_3$ & 21$_{2,20}-$20$_{2,19}$(EE) & 122 & $-$\\
219.276 & SO$_2$ & 22$_{7,15}-$23$_{6,18}$ & 353 & 1.96(06)\\
219.355 & $^{34}$SO$_2$ & 11$_{1,11}-$10$_{0,10}$ & 60 & $-$\\
219.401 & C$_2$H$_3$CN & 23$_{3,20}-$22$_{3,19}$ & 145 & $-$\\
219.466 & SO$_2$;$v_2$=1 & 22$_{2,20}-$22$_{1,21}$ & 1013 & $-$\\
219.506 & C$_2$H$_5$CN & 24$_{2,22}-$23$_{2,21}$ & 136 & $-$\\
219.560 & C$^{18}$O & 2$-$1 & 16 & 1.18(04)\\
219.616 & unidentified & && \\
219.657 & HNCO & 10$_{3,8}-$9$_{3,7}$ & 433 & $-$\\
 & HNCO & 10$_{3,7}-$9$_{3,6}$ & 433 & $-$\\
219.675 & HC$_3$N;$v_7$=2 & 24$-$23,l=0 & 773 & $-$\\
219.707 & HC$_3$N;$v_7$=2 & 24$-$23,l=2e & 777 & $-$\\
219.734 & HNCO & 10$_{2,9}-$9$_{2,8}$ & 228 & $-$\\
 & HNCO & 10$_{2,8}-$9$_{2,7}$ & 228 & $-$\\
 & HC$_3$N;$v_7$=2 & 24$-$23,l=2f & 777 & $-$\\
219.798 & HNCO & 10$_{0,10}-$9$_{0,9}$ & 58 & 6.86(06)\\
219.909 & H$_2^{13}$CO & 3$_{1,2}-$2$_{1,1}$ & 33 & $-$\\
219.949 & SO & 6$_{5}-$5$_{4}$ & 35 & 2.27(06)\\
219.994 & CH$_3$OH & 23$_{5,18}-$22$_{6,17}$(E) & 776 & $-$\\
220.038 & t-HCOOH & 10$_{0,10}-$9$_{0,9}$ & 59 & $-$\\
220.079 & CH$_3$OH & 8$_{0,8}-$7$_{1,6}$(E) & 97 & 3.40(07)\\
 & HC$_3$N;$v_5$=1/$v_7$=3 & 24$-$23,$v_7$=3$^{1-}$ & 1087 & $-$\\
220.165 & SO$_2$;$v_2$=1 & 16$_{3,13}-$16$_{2,14}$ & 910 & $-$\\
 & CH$_3$OCHO & 17$_{4,13}-$16$_{4,12}$(E) & 103 & $-$\\
220.178 & CH$_2$CO & 11$_{1,11}-$10$_{1,10}$ & 76 & $-$\\
220.190 & CH$_3$OCHO & 17$_{4,13}-$16$_{4,12}$(A) & 103 & $-$\\
220.355 & CH$_3$COCH$_3$ & 22$_{0,22}-$21$_{0,21}$(AE\&EA) & 124 & $-$\\
 & CH$_3$COCH$_3$ & 22$_{1,22}-$21$_{1,21}$(EA) & 124 & $-$\\
220.362 & CH$_3$COCH$_3$ & 22$_{1,22}-$21$_{1,21}$(EE) & 124 & $-$\\
 & CH$_3$COCH$_3$ & 22$_{0,22}-$21$_{0,21}$(EE) & 124 & $-$\\
220.368 & CH$_3$COCH$_3$ & 22$_{1,22}-$21$_{1,21}$(AA) & 124 & $-$\\
 & CH$_3$COCH$_3$ & 22$_{0,22}-$21$_{0,21}$(AA) & 124 & $-$\\
220.399 & $^{13}$CO & 2$-$1 & 16 & 1.18(04)\\
 & CH$_3$OH & 10$_{-5,6}-$11$_{-4,8}$(E) & 252 & 9.32(06)\\
220.407 & HC$_3$N;$v_5$=1/$v_7$=3 & 24$-$23,$v_7$=3$^{3-}$ & 1094 & $-$\\
 & HC$_3$N;$v_5$=1/$v_7$=3 & 24$-$23,$v_7$=3$^{3+}$ & 1094 & $-$\\
220.476 & CH$_3$CN & 12$_{8}-$11$_{8}$ & 526 & 2.36(06)\\
220.539 & CH$_3$CN & 12$_{7}-$11$_{7}$ & 419 & 2.71(06)\\
220.585 & HNCO & 10$_{1,9}-$9$_{1,8}$ & 102 & 7.95(06)\\
220.594 & CH$_3$CN & 12$_{6}-$11$_{6}$ & 326 & 3.44(06)\\
 & CH$_{3}^{13}$CN & 12$_{3}-$11$_{3}$ & 133 & $-$\\
220.618 & $^{33}$SO$_2$ & 11$_{1,11}-$10$_{0,10}$ & 61 & $-$\\
 & CH$_{3}^{13}$CN & 12$_{2}-$11$_{2}$ & 97 & $-$\\
220.641 & CH$_3$CN & 12$_{5}-$11$_{5}$ & 247 & 3.41(06)\\
 & CH$_{3}^{13}$CN & 12$_{1}-$11$_{1}$ & 76 & $-$\\
 & CH$_{3}^{13}$CN & 12$_{0}-$11$_{0}$ & 69 & $-$\\
220.658 & SO$_2$;$v_2$=1 & 12$_{5,7}-$13$_{4,10}$ & 896 & $-$\\
 & C$_2$H$_5$CN & 25$_{2,24}-$24$_{2,23}$ & 143 & $-$\\
220.679 & CH$_3$CN & 12$_{4}-$11$_{4}$ & 183 & 3.73(06)\\
220.709 & CH$_3$CN & 12$_{3}-$11$_{3}$ & 133 & 4.05(06)\\
220.730 & CH$_3$CN & 12$_{2}-$11$_{2}$ & 97 & 3.99(06)\\
220.743 & CH$_3$CN & 12$_{1}-$11$_{1}$ & 76 & 3.92(06)\\
 & CH$_3$CN & 12$_{0}-$11$_{0}$ & 69 & 4.23(06)\\
\hline
\end{supertabular}
\tablefoot{a(b) = a$\times$10$^{\mathrm{b}}$. The collisional rate coefficients measured at the following temperatures were used to calculate the critical density: $^{(a)}$ $C_{\mathrm{ul}} (T = 200$\,K). $^{(b)}$ $C_{\mathrm{ul}} (T = 160$\,K). $^{(c)}$ $C_{\mathrm{ul}} (T = 140$\,K).}
\end{center}
\end{scriptsize}
Depending on the nuclear spin states of the three hydrogen atoms in the methyl group, CH$_3$OH and CH$_3$OCHO are denoted with A (parallel proton spin) or E (anti-parallel proton spin), whereas CH$_3$OCH$_3$ and CH$_3$COCH$_3$ have two methyl groups (denoted with AA, EE, AE, or EA). Formic acid (HCOOH) exists in two conformers (c-HCOOH and t-HCOOH) depending on the orientation of the hydroxy (OH) group. Multiple transitions at the same frequency are blended within our spectral setup at a spectral resolution of 2.2\,MHz.
\section{Spectral line modeling with {\tt XCLASS}}\label{section:linefitting}
\begin{figure*}
\centering
\includegraphics[height=15.5cm, angle=90]{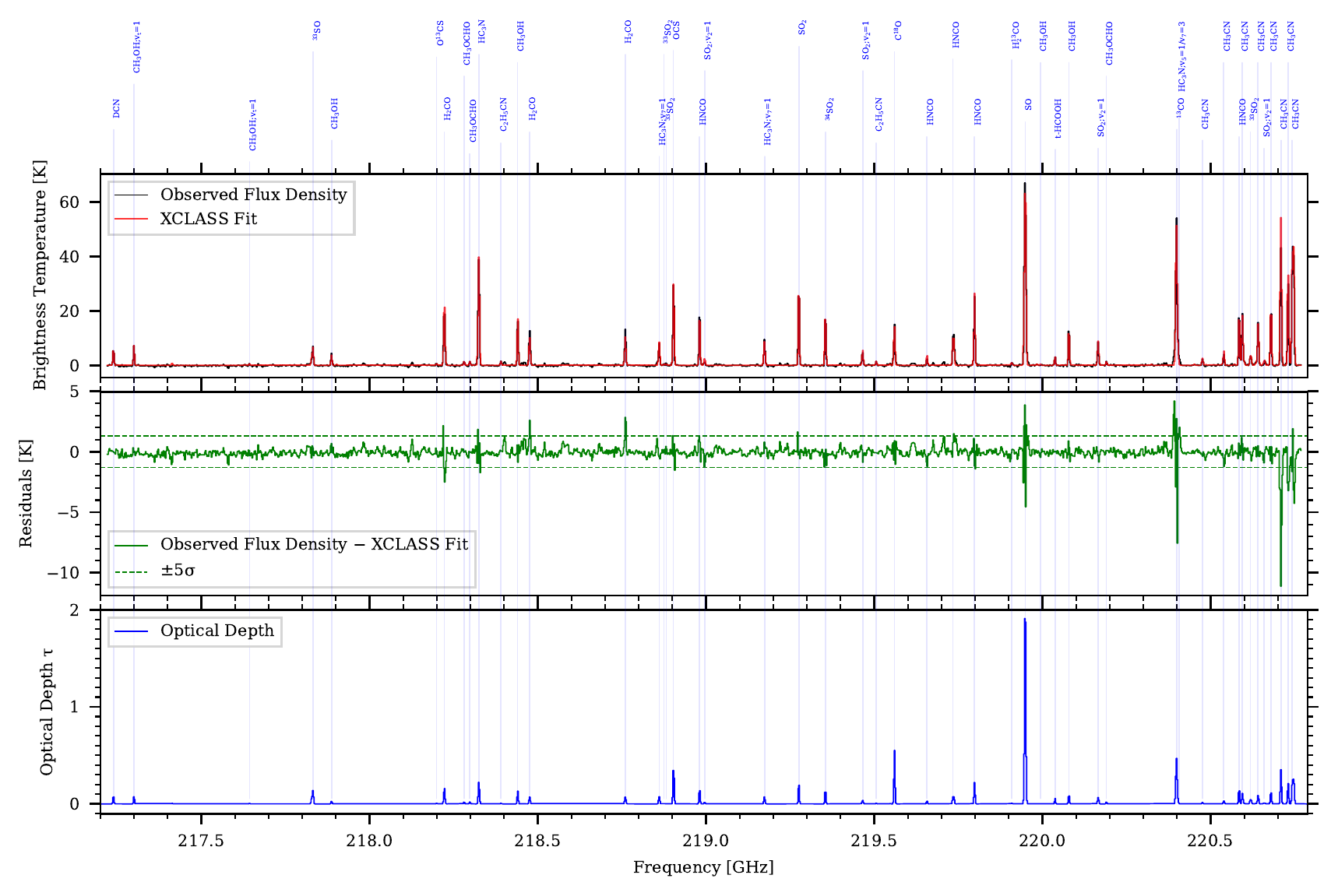}
\caption{Comparison of the observed and fitted {\tt XCLASS} spectrum. Upper panel: AFGL\,2591 spectrum towards the continuum peak (black line) and the total {\tt XCLASS} fit spectrum (red line). Middle panel: Residuals (green line). Bottom panel: Optical depth $\tau$ (blue line) of the transitions computed with {\tt XCLASS}.}
\label{Fig:XCLASSFit}
\end{figure*}

Similar to the map fitting described in Sect. \ref{subsec:T_profile}, we use {\tt XCLASS} to fit the spectrum extracted towards the position of the 1.37\,mm continuum peak. In order to find the best-fit parameters of all detected molecules with {\tt XCLASS}, an algorithm chain with the Genetic (50 iterations) and Levenberg–Marquardt algorithm (50 iterations) is adopted \citep[a detailed description of the algorithms is given in][]{MAGIX}. In the spectrum we estimate a rms noise of 0.26\,K in a line-free frequency range at $217.664 - 217.706$\,GHz. For each molecule that is detected with emission lines $S/N > 5$, the {\tt myXCLASSFit} function models the non-blended emission lines with one emission component in order to find the best-fit parameters. In order to estimate the uncertainties of the fit parameters, we scale the intensity of spectrum to both $+20$\,\% and $-20$\,\%, which is the maximum expected uncertainty from the flux calibration of the CORE data, as discussed in \citet{Beuther2018}. The two scaled spectra are likewise fitted with {\tt myXCLASSFit} and the uncertainties are then estimated by the mean deviation from the best-fit parameters. A comparison between the observed and fitted spectrum is shown in Fig. \ref{Fig:XCLASSFit}. There are large differences between the observed and modeled emission lines for molecules which have either non-Gaussian line profiles and/or a high optical depth (e.g., $^{13}$CO, SO, H$_2$CO and the CH$_3$CN $K = 3$ transition). 

The underlying assumption in the calculation of the synthetic spectra with {\tt XCLASS} is that LTE conditions hold. In order to check that this assumption is valid, we use the non-LTE radiative transfer code {\tt RADEX}\footnote{\url{http://var.sron.nl/radex/radex.php}} \citep{RADEX} to calculate the rotation temperature $T_{\mathrm{rot}}$ of the molecules with the derived physical parameters as an input. We assume the CMB (cosmic microwave background) temperature for the background temperature ($T_{\mathrm{bg}} = 2.73$\,K). For the kinetic temperature, we use the CH$_3$CN temperature $T_{\mathrm{kin}} = 200$\,K and column densities of each molecule derived towards the continuum peak. We assume the H$_2$ volume density of n(H$_2) \approx 10^7$\,cm$^{-3}$ described in Sect. \ref{subsec:h2gas} and the linewidth of $\Delta v = 5.2$\,km\,s$^{-1}$ described in Sect. \ref{sec:XCLASSFittPeak}. Local thermal equilibrium conditions are given when $T_{\mathrm{rot}} = T_{\mathrm{kin}}$.

\begin{table}[htb]
\centering
\caption{Rotation temperatures calculated with {\tt RADEX}.}
\label{tab:radex}
\setlength{\tabcolsep}{5pt}
\begin{tabular}{lc|lc}
\hline \hline
Species (Transition) & $T_{\mathrm{rot}}$ & Species (Transition) & $T_{\mathrm{rot}}$\\
& (K) & & (K) \\
\hline
$^{13}$CO ($2-1$) & 200 & CH$_3$OH (4$_{2,3}-3_{1,2}$) & $-$40 \\
C$^{18}$O ($2-1$) & 201 & CH$_3$OH (8$_{0,8}-7_{1,6}$)& $-$85 \\
SO (6$_{5}-5_{4}$) & 206 & CH$_3$OH (10$_{-5,6}-11_{-4,8}$)& 10\\
SO$_2$ (22$_{7,15}-23_{6,18}$)& 9 & CH$_3$CN (12$_{6}-11_{6}$) & 152 \\
OCS ($18-17$) &198& CH$_3$CN (12$_{5}-11_{5}$) & 178\\
H$_2$CO (3$_{0,3}-2_{0,2}$) & 229 &CH$_3$CN (12$_{4}-11_{4}$) & 164 \\
H$_2$CO (3$_{2,2}-2_{2,1}$)& 269 & HNCO(10$_{1,10}-9_{1,9}$) &108 \\
H$_2$CO (3$_{2,1}-2_{2,0}$)& 253 & HNCO (10$_{0,10}-9_{0,9}$)& 191 \\
 & & HNCO (10$_{1,9}-9_{1,8}$) & 147 \\
\hline
\end{tabular}
\end{table}

Table \ref{tab:radex} shows the computed rotation temperature for all molecular column densities towards the continuum peak that have available data in {\tt RADEX}. For most molecules $T_{\mathrm{ex}} \approx T_{\mathrm{kin}}$, except for SO$_2$ (9\,K) and CH$_3$OH (10\,K). As vibrational excited SO$_2$ with high upper energy levels ($\gtrsim 900$\,K) is also detected, part of the emission has to stem from a non-LTE environment \citep[e.g., in a shocked region as studied by][]{Jorgensen2004}. The negative rotation temperatures of CH$_3$OH denote that maser activity is expected and indeed hot cores are known to show strong CH$_3$OH maser emission \citep[e.g.,][]{Longmore2007}.
\end{appendix}
\end{document}